\documentclass[reqno,12pt]{article}
\usepackage{ae} 
\usepackage[T1]{fontenc}
\usepackage[ansinew]{inputenc}
\usepackage{amsmath}
\usepackage{amsfonts}
\usepackage{amssymb}
\usepackage{graphicx}
\usepackage{bm}
\usepackage{float}
\usepackage{color}
\definecolor{darkblue}{cmyk}{0.9,0.9,0,0}
\usepackage[colorlinks=true,linkcolor=darkblue,citecolor=darkblue,urlcolor=darkblue]{hyperref}
\usepackage{epsfig}

\usepackage[dvipsnames]{xcolor}
\usepackage{graphicx}

\usepackage{diagbox}

\newcommand{\beq}{\begin{equation}}
\newcommand{\eeq}{\end{equation}}
\newcommand\beqa{\begin{eqnarray}}
\newcommand\eeqa{\end{eqnarray}}
\newcommand\bea{\begin{array}}
\newcommand\eea{\end{array}}

\newcommand{\Li}{\text{Li}}

\def\XXint#1#2#3{{\setbox0=\hbox{$#1{#2#3}{\int}$}
\vcenter{\hbox{$#2#3$}}\kern-.5\wd0}}

\newcommand{\nn}{\nonumber}

\newcommand{\neqa}{\nonumber\end{eqnarray}}
\newcommand{\la}[1]{\label{#1}}

\newcommand{\Tr}{{\rm Tr}}

\renewcommand{\d}{\partial}

\newcommand{\<}{{\langle}}
\renewcommand{\>}{{\rangle}}

\newcommand{\re}{\relax{\rm I\kern-.18em R}}

\renewcommand{\sp}{p\hspace{-.40em}/}

\def\su2{{SU(2)}}

\def\[{\left[}
\def\]{\right]}

\def\({\left(}
\def\){\right)}
\def\[{\left[}
\def\]{\right]}

\def\<{\langle}
\def\>{\rangle}

\def\i2{\frac{i}{2}}

\def\spi{\relax{\rm \pi\kern-0.5em /}}
\def\sA{\relax{\rm A\kern-0.5em /}}
\def\sp{\relax{\rm p\kern-0.5em /}}
\def\sd{\relax{\rm \d\kern-0.5em /}}
\def\sk{\relax{\rm k\kern-0.5em /}}
\def\sn{\relax{\rm n\kern-0.5em /}}
\def\sl{\relax{\rm l\kern-0.5em /}}
\def\sP{\relax{\rm P\kern-0.7em /}}
\def\sBethe{\relax{\rm \Bethe\kern-0.5em /}}

\usepackage{varioref}
\usepackage{makeidx}
\makeindex

\newcommand\blfootnote[1]{%
  \begingroup
  \renewcommand\thefootnote{}\footnote{\hspace{-6mm}#1}%
  \addtocounter{footnote}{-1}%
  \endgroup
}

\usepackage[english]{babel}
\begin{document}


\thispagestyle{empty}

\renewcommand{\thefootnote}{\fnsymbol{footnote}}
\setcounter{page}{1}
\setcounter{footnote}{0}
\setcounter{figure}{0}

\vspace{-0.4in}

\begin{center}
$$$$
{\Large\textbf{\mathversion{bold}
Stampedes I:\\
Fishnet OPE and Octagon Bootstrap with Nonzero Bridges
}\par}
\vspace{1.0cm}

\textrm{Enrico Olivucci$^\text{\tiny 1}$, Pedro Vieira$^\text{\tiny 1,\tiny 2}$}
\blfootnote{\tt  \#@gmail.com\&/@\{e.olivucci,pedrogvieira\}}
\\ \vspace{1.2cm}
\footnotesize{\textit{
$^\text{\tiny 1}$Perimeter Institute for Theoretical Physics,
Waterloo, Ontario N2L 2Y5, Canada \\
$^\text{\tiny 2}$ICTP South American Institute for Fundamental Research, IFT-UNESP, S\~ao Paulo, SP Brazil 01440-070   \\
}  
\vspace{4mm}
}
\end{center}

\par\vspace{1.5cm}


\vspace{2mm}
\begin{abstract}
Some quantities in quantum field theory are dominated by so-called \textit{leading logs} and can be re-summed to all loop orders. In this work we introduce a notion of \textit{stampede} which is a simple time-evolution of a bunch of particles which start their life in a corner -- on the very right say -- and \textit{hop} their way to the opposite corner -- on the left -- through the repeated action of a quantum Hamiltonian. Such stampedes govern leading logs quantities in certain quantum field theories. The leading euclidean OPE limit of correlation functions in the fishnet theory and null double-scaling limits of correlators in $\mathcal{N}=4$ SYM are notable examples. As an application, we use these results to extend the beautiful bootstrap program of Coronado \cite{Frank} to all octagons functions with arbitrary diagonal bridge length.

\end{abstract}

\newpage

\setcounter{page}{1}
\renewcommand{\thefootnote}{\arabic{footnote}}
\setcounter{footnote}{0}



{
\tableofcontents
}



\newpage

\section{Introduction} 
\label{sect:intro}

\begin{figure}[t]
\begin{center}
{\includegraphics[scale=0.38]{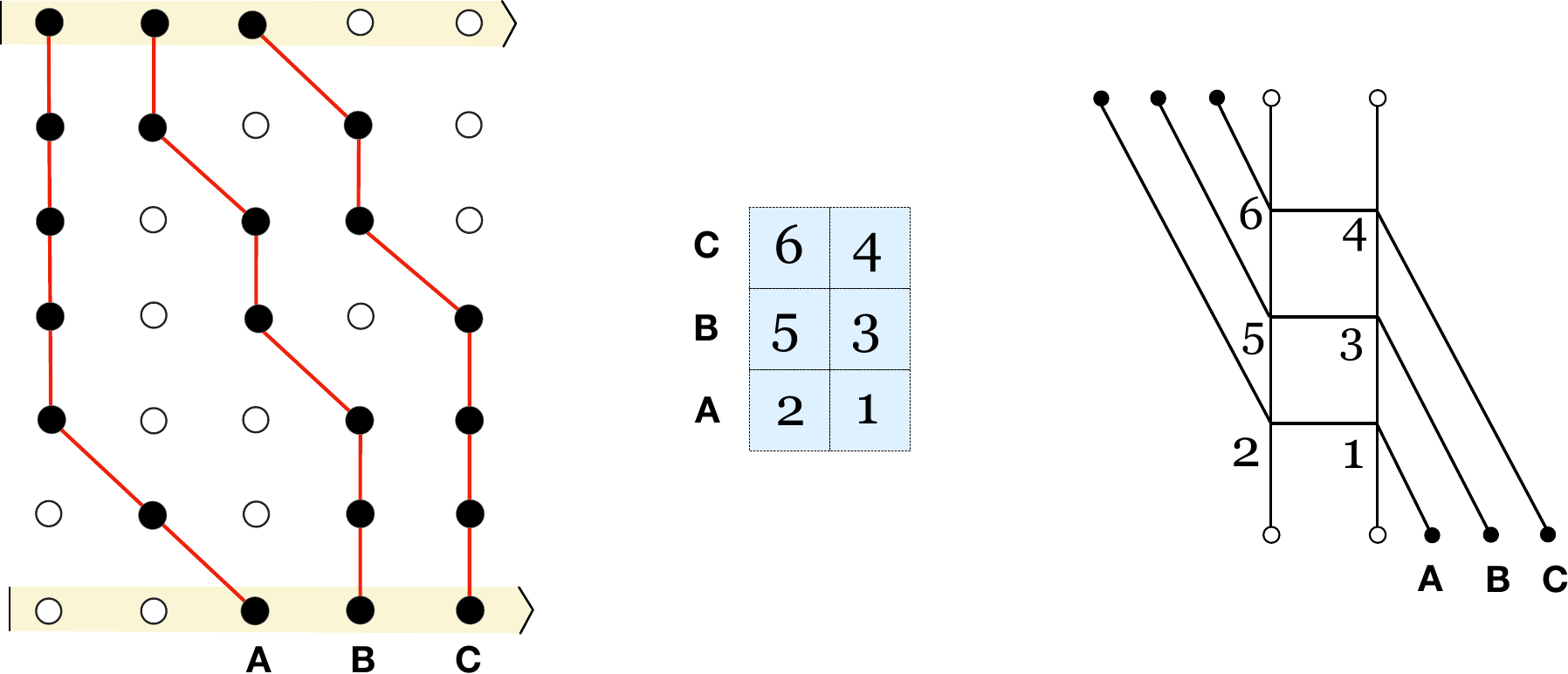}}
\end{center}
\label{SYT_BD}
\caption{\textit{Left:} one possible stampede of $m=3$ magnons with $n=2$ empty sites. The initial state (bottom) is $|\circ \circ \bullet \bullet\, \bullet \rangle $, the final state (top) is $| \bullet \bullet \bullet \circ \, \circ \rangle$. Red lines are the trajectories of the magnons. \textit{Center:} Young tableaux associated with the stampede. \textit{Right:} Graph associated to the stampede: when one magnon $A,B,C$ crosses a vertical line, it hops one step left. The vertices are ordered according to the stampede on the left picture. }
\end{figure}

In quantum spin chains magnons \textit{hop}. That is, by action of the spin chain Hamiltonian, magnon excitations move, typically to its neighbouring site to the left or to the right. We can imagine several magnon excitations running from a right region of a quantum spin chain to a left region. We call it a \textit{stampede}. Quantum mechanically there are many possible paths for the particles to move from the right to the left and the stampede is the sum over all such possibilities as usual. 

As a first example, consider a bunch of magnon excitations in a \textit{chiral nearest neighbour Hamiltonian} where magnons can hop to the left but not to the right. That is, 
\beq
H |\dots \circ\circ\circ\bullet \circ\circ\dots \rangle= g |\dots \circ\circ\overset{{\color{gray}\curvearrowleft}}{\bullet\,\circ} \circ\circ\dots \rangle \,
\eeq
(we added an arrow in light grey to keep track of what just happened). 
The magnons are hardcore so only one magnon ($\bullet$) can occupy a spin chain site which can otherwise be empty~$(\circ)$\footnote{We can think of $\circ$ as a spin down and $\bullet$ be a spin up. Then the $\bullet$ excitations would be magnon fluctuations of a ferromagnetic vacua made out of $\circ$ only.}. For instance, we have 
\beq
H |\dots \circ\circ\circ\bullet\bullet \circ\circ\dots \rangle= g |\dots \circ\circ \overset{{\color{gray}\curvearrowleft}}{\bullet\,\circ}  \bullet \circ\circ\dots \rangle\,
\eeq
since only the left magnon can move; the right magnon can not move because the left magnon is occupying a space. Of course, if we act twice with the Hamiltonian in this example we have two possibilities: the left particle can keep moving to the left \textit{or} the right magnon can now move to the left also since there is now room, 
\beq
H^2 |\dots \circ\circ\circ\bullet\bullet \circ\circ\dots \rangle= g^2 |\dots \circ \overset{{\color{gray}\curvearrowleft}}{\bullet\,\circ} \circ\bullet \circ\circ\dots \rangle + g^2 |\dots \circ\circ\bullet\overset{{\color{gray}\curvearrowleft}}{\bullet\,\circ} \circ\circ\dots \rangle\,. 
\eeq
We can now define the stampede. We start with $m$ magnons at the very right and $n$ empty sites at the left. After $nm$ actions of the chiral Hamiltonian all these magnons will have moved to the very left,
\beq
H^{nm} | \underbrace{\circ\dots \circ}_{n} \underbrace{\bullet \dots \bullet}_{m}  \rangle= g^{nm} \texttt{(fishnet stampede)}_{n,m} | \underbrace{\bullet \dots \bullet}_{m}  \underbrace{\circ\dots \circ}_{n}  \rangle\,,
\eeq
where $ \texttt{(fishnet stampede)}_{n,m}$ is the number of ways such hardcore magnons can reach the left starting from the right in $nm$ steps, see figure \ref{SYT_BD}. The reason for the \texttt{fishnet} qualifier will be clear soon. 
The number $ \texttt{(fishnet stampede)}_{n,m}$  is known in a combinatorics context as the solution of the \emph{ballot problem} \cite{Frobenius,Hook,Gessel}
\beq
\label{Catalan_mn}
\texttt{(fishnet stampede)}_{n,m} = \frac{(n m)!}{\prod_{i=1}^m\prod_{j=1}^n (m+n-i-j+1)}\,.
\eeq
Details on the derivation of \eqref{Catalan_mn} are presented in the appendix \ref{app:Catalan}. In synthesis, the number of possible stampedes of $m$ hard magnons that all hop left by $n$ steps is equal to the possible fillings of a $m\times n$ table with numbers $1,2,\dots, m n$ that strictly decrease along each column/row. The $k$-th row is associated with the $k$-th magnon from the right, and it gets filled with the numbers $1\leq k_1,\dots,k_n\leq mn$ of the steps when that magnon hops. Such tables are rectangular Young diagrams and the fillings are the associated standard Young tableaux.
We present an example for $m=3$ and $n=2$ in figure \ref{SYT_BD}.

\begin{figure}[t]
 \begin{center}
{\includegraphics[scale=0.75]{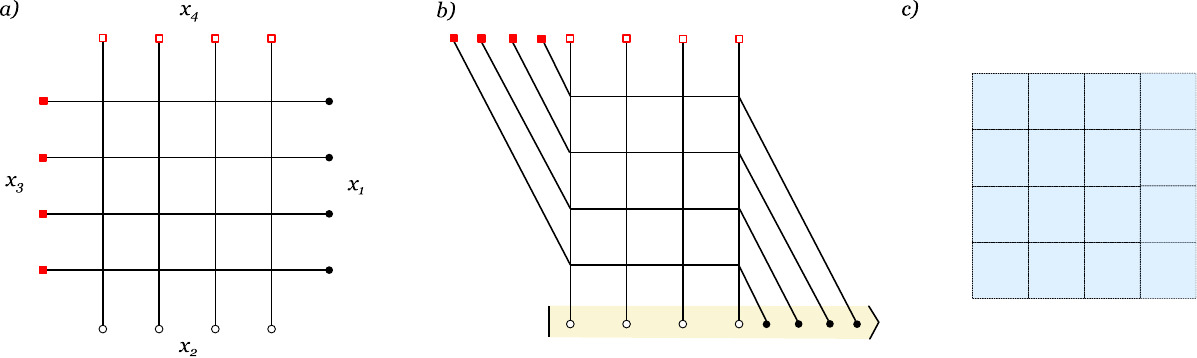}}
\end{center}
\caption{Fishnet Feynman integrals of rectangular shape and the stampede. The symbols $\bullet,\circ$ denote the external fields $X,Z$, while $\color{red}{ \blacksquare} \color{black},\color{red}\square $ are $\bar X,\bar Z$ respectively. Each crossing of lines is an integration point (quartic vertex) with coupling $g$. Fields of the same type lie at the same space-time point.  \textit{a)} Rectangular fishnet of size $m\times n = 4\times 4$. \textit{b)} The limit where the position of fields $X,{Z}$ coincide, and the same for $\bar Z,\bar X$. We highlight the initial ket in the corresponding stampede process. \text{c)} Young diagram associated with the graph by replacement of each quartic vertex with a box.}
\label{BD}
 \end{figure}

It turns out that $ \texttt{(fishnet stampede)}_{n,m}$ nicely governs the leading OPE behaviour of a general class of $4D$ planar Feynman diagrams introduced by Basso and Dixon in \cite{BassoDixon} and recently explored in several limits -- including the OPE limit -- in \cite{BassoDixonAndFriends}. Referring to figure \ref{BD}, the basic idea is to regard the $m$ fields $X$ emitted in $x_1$ as magnons ($\bullet$) and the fields $Z$ emitted in $x_2$ as empty spaces ($\circ$).
 \noindent
Each of the $m$ magnons $X$ propagates from $x_1$ to $x_3$, crossing all of the $n$ propagators of empty spaces $Z$ that connect $x_2$ to $x_4$, and giving rise to a rectangular square lattice of $mn $ quartic vertices $XZ\bar X \bar Z$, showed in figure \ref{BD}a. The absence of any other vertices means we are dealing with \emph{chiral hardcore} magnons which hop only left. Therefore, the OPE limit of the correlator $x_2\to x_1,\,x_4\to x_3$ is described by the evolution of a state $| {\circ\dots \circ} \,{\bullet \dots \bullet}  \rangle$ in $x_1$ to a state $|{\bullet \dots \bullet} \,{\circ\dots \circ}   \rangle$ in $x_3$ by the chiral fishnet Hamiltonian~\cite{FishnetJoao} as in figure \ref{BD}b. The trajectory followed by the magnons is not specified, therefore they all contribute and their number multiplies the leading logarithm
\beq
\label{BD_stamp_intro}
\Phi_{nm}=\frac{(-\log u)^{nm}}{(nm)!} \times \texttt{(fishnet stampede)}_{n,m} \,,
\eeq
where we introduced the conformal cross-ratio $u = (x_{12}^2 x_{34}^2)/(x_{14}^2x_{23}^2)$. 
This is precisely formula (4.20) from \cite{BassoDixonAndFriends}. 

With this simple chiral stampede appetizer we close the introduction. In the rest of the paper we consider richer stampedes. In the next section we consider chiral stampedes with generalized initial and final end-points and even with particles leaving and entering the race mid stamped thus generating several new predictions for more general fishnet correlators. Next we move to more general non-chiral stampedes where particles can move to the left and to the right, they can form groups on the same site and they can even jump in groups when hopping from one site to the neighbour. Such stampedes arise from Hamiltonians of non-compact spin chains. They arise in large $N$ gauge theories when studying spinning operators where gluons dominate. We will see that such stampedes thus govern the light-cone kinematics of several correlation functions in some double-scaling limits where the coupling is sent to zero at the same time as points are sent to null configurations. They will allow us to extend Coronado's bootstrap \cite{Frank, FrankBootstrap} of the simplest correlators in $\mathcal{N}=4$ SYM to a larger family of correlators.

\begin{figure}[t]
 \begin{center}
{\includegraphics[scale=0.7]{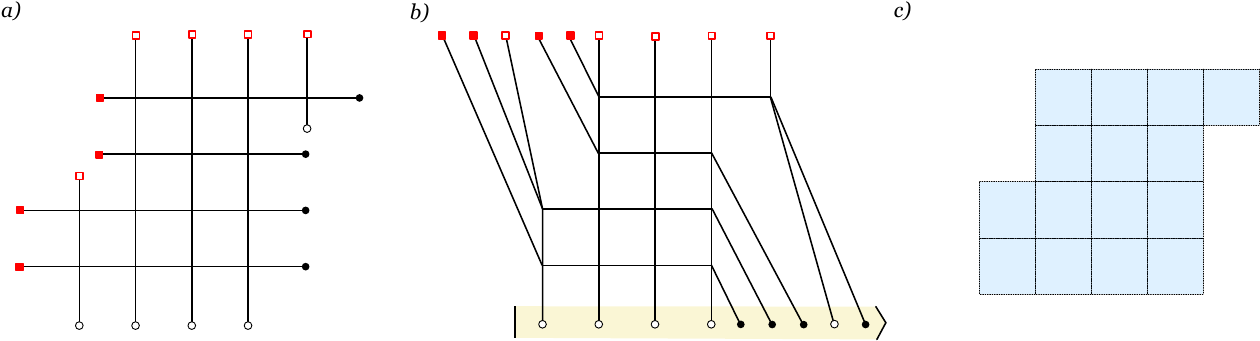}}
\end{center}
\caption{Fishnet integrals of skew shape $\lambda/\mu$. Fields of the same type that lie on a same vertical or horizontal segment are set at the same space-time point.  \textit{a)} Integral with skew shape $\lambda =\{5,4,4,4\}\,,\, \mu =\{1,1,0,0\}$, appearing at order $g^{15}$ in Fishnet CFT. \textit{b)} The limit where the position of fields $X,{Z}$ coincide, and the same for $\bar Z,\bar X$. We highlight the initial ket of the stampede process. \text{c)} Skew Young diagram associated with the graph.}\label{Skew}
 \end{figure}
 \noindent

\section{General stampedes of hard magnons}

\label{sect:fishnet}
The rectangular Basso-Dixon fishnets studied in the previous section appear in the context of Fishnet CFT \cite{Fishnet,FishnetJoao} as the only Feynman diagrams contributing to the perturbation theory of single-trace correlators of the type
\begin{equation}
\label{BD_correlator}
I_{m,n}^{BD}(x_1,x_2,x_3,x_4) = \langle \text{Tr} \left[X^m(x_1) Z^n(x_2)\bar X^m (x_3) \bar Z^n (x_4) \right]\rangle\,.
\end{equation}
Such a simple description of the correlator is due to the chirality of the Fishnet action, namely the fact that the only allowed interaction $\Tr\left(XZ\bar X \bar Z\right)$ is not self-adjoint. In fact, the non-unitarity of the theory explains why the OPE of $I_{m,n}^{BD}$ truncates at order $\log(z\bar z)^{mn}$, according to \eqref{BD_stamp_intro}.
\begin{figure}[t]
\begin{center}
{\includegraphics[scale=0.35]{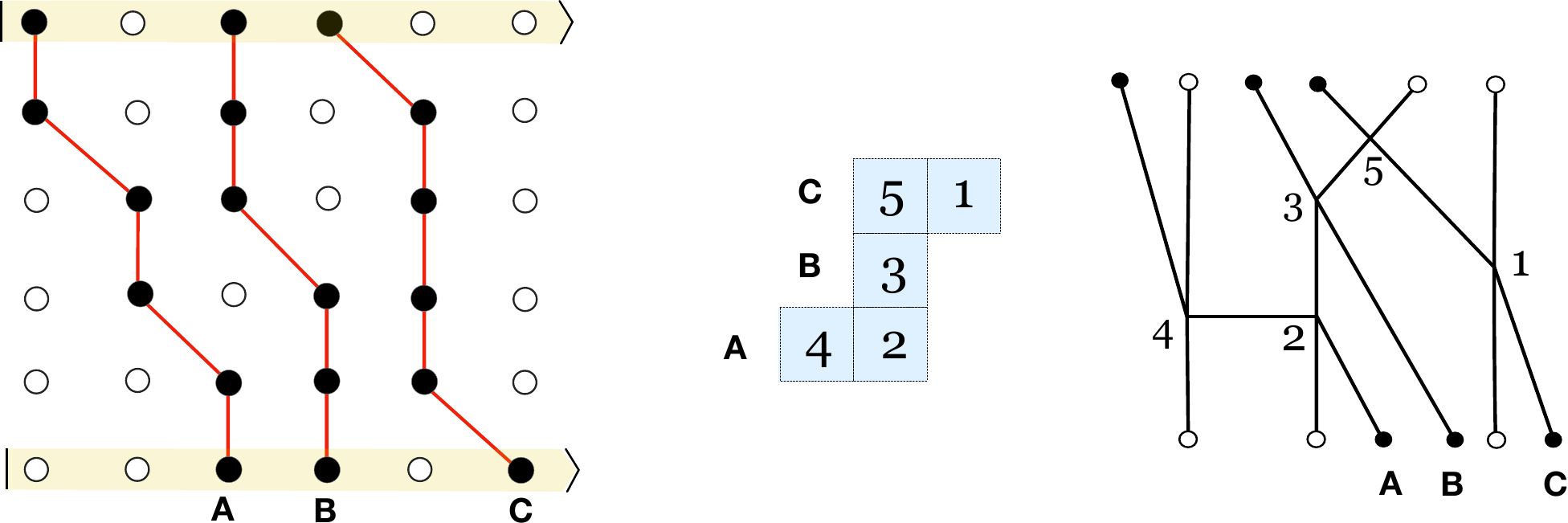}}
\end{center}
\caption{\textit{Left}: A possible way of hopping of $m=3$ hard magnons - labelled as $A,B,C$ - from a certain initial to final state with empty spaces among magnons. \textit{Center}: The standard skew Young tableaux corresponding to this process. \textit{Right}: Graph representation of the process: each magnon $A,B,C$ hops left when crossing a vertical line. Vertices are numbered according to the order of hoppings.}
\label{Skew_hop1}
\end{figure}
\noindent

\begin{figure}[t]
 \begin{center}
{\includegraphics[scale=0.7]{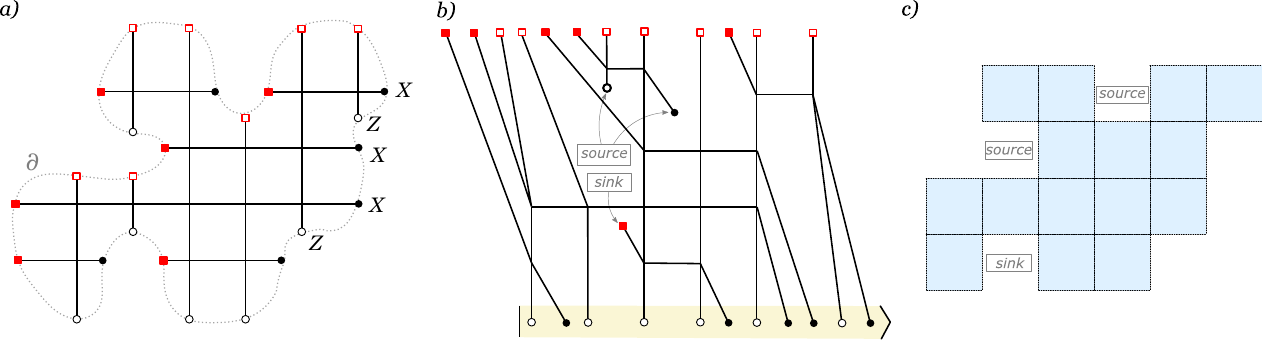}}
\end{center}
\caption{Fishnet on the disk with boundary $\partial=\{XZXXZ\dots\}$. Fields of the same type that lie on a same vertical or horizontal segment lie at the same space-time point.  \textit{a)} Example of a $20$-point integral at order $g^{15}$ in Fishnet CFT. \textit{b)} The limit where the position of fields $X,{Z}$ coincide, and the same for $\bar Z,\bar X$. We highlight the \emph{sources} and \emph{sinks} of magnons, corresponding to concavities of the diagram. \text{c)} Generalized Young diagram associated with the graph.}
\label{Gen}
 \end{figure}

Formula \eqref{BD_stamp_intro} can be extended to the leading OPE of the general higher-point single-trace correlator in Fishnet theory. These differ from \eqref{BD_correlator} by the fact that the fields $X,Z,\bar X,\bar Z$ are arranged in more generic way inside the trace, corresponding to a specific boundary condition~$\partial$ on the fishnet square-lattice.
We distinguish two cases by increasing generality, exemplified by figures \ref{Skew}a and \ref{Gen}a.
In the fishnet integrals of the type \ref{Skew}a, the $j$-th horizontal line crosses the last $(\lambda_j-\mu_j)$ vertical lines, where $\lambda_i\leq \lambda_{i-1},\,\mu_i \leq \mu_{i-1}$. Such integrals describe single-trace correlators between $m= \text{length}{({\lambda})},\,n={\lambda_1}$ fields of type $X,Z$ (and their conjugates) respectively. 
For instance, in figure \ref{Skew}a we have
\beq
 \lambda =\{5,4,4,4\}\,,\,\,\,\,\, \mu =\{1,1,0,0\}\,,
\eeq
and the single-trace correlator is
\beq
\label{skew_corr_fish}
I_{{\lambda}/{\mu}}(x_1,\dots,x_8) = \langle \text{Tr} \left[X (x_1)Z(x_2) X^3 (x_3) Z^4(x_4) \bar X^2 (x_5) \bar Z(x_6) \bar X^2 (x_7) \bar Z^4(x_8) \right]\rangle\,.
\eeq
The coefficient of the leading $\log$ in the limit where fields $X,Z$ and $\bar X,\bar Z$ are pinched respectively at two space-time points is captured by the chiral stampede with initial and final states where the magnons are not all adjacent, as illustrated in figure \ref{Skew}b. Each of these integrals can be mapped to a skew Young diagram of shape ${\lambda}/{\mu}$ by the replacement of each quartic vertex with a box (figure \ref{Skew}c).

The counting of the possible stampedes is equal to the number of standard skew Young tableaux of shape $\lambda/\mu$, namely the possible filling of the skew Young diagram with strictly decreasing integers $\{1,\dots,\# \text{boxes}\}$ along its rows/columns, as in the example of figure \ref{Skew_hop1}. 

Following the combinatoric analysis of \cite{Gessel}, a planar fishnet with $L = \sum_j ({\lambda_j-\mu_j})$ vertices and shape $\lambda/\mu$, we conjecture that the OPE leading logarithm is
\beq
\Phi_{\lambda /\mu}=\frac{(-\log u)^{L}}{L!} \times\texttt{(fishnet stampede)}_{\lambda/\mu}  \,,
\eeq
for $u$ a cross-ratio of \eqref{skew_corr_fish} such that $u \to 0$ in the the OPE limit. 
The explicit computation is given in the form of determinants by the Aitken's formula \cite{Aitken,Hook},
\beq
\texttt{(fishnet stampede)}_{\lambda/\mu} = \left(\sum_{j=1}^{m} \left(\lambda_j-\mu_j\right) \right) ! \, \det_{1\leq i,j\leq m}\left(\frac{1}{(\lambda_i-i-\mu_j+j)!}\right)\,,
\eeq
and for a rectangular shape $\lambda_i=\lambda_j$, $\mu_i=0$ it reduces to the $\texttt{(fishnet stampede)}_{n,m}$ treated in the first section. For example, the leading $\log$ of the correlator \eqref{skew_corr_fish} reads
\beq
\Phi_{\{5,4,4,4\}/\{1,1,0,0\}} = \frac{(-\log u)^{15}}{15!}\times 99099 \,.
\eeq
\noindent
The figure \ref{Gen}a provides an example of a very general fishnet integral on the disk \cite{gen_fish}, that is a portion of square lattice that can be cut out a plane by drawing a closed contour $\partial$. The single-trace higher-point Fishnet correlator corresponding to \ref{Gen}a is defined by the contour $\partial$ that correspond to a certain permutation of fields $X^mZ^n\bar X^m \bar Z^n$ inside the trace. For example, in figure \ref{Gen} starting from the up-right field $\bullet=X(x_1)$ and proceeding clockwise we have
\beq
\medmuskip=-1mu
\thinmuskip=1mu
\thickmuskip=-1mu
I_{\partial} = \langle\Tr\left[ X (x_1)  Z(x_2) X^2(x_3) Z(x_4) X(x_5)Z^2(x_6) \bar X(x_7)Z(x_8) X(x_9)Z(x_{10}) \bar X^2(x_{12})\cdots \bar{Z}^2(x_{20})\right] \rangle\,. 
\eeq
In this case the hopping process that captures the OPE leading log coefficient is dynamic, as the number of magnons can change during the stampede, in correspondence of concavities of the fishnet shape (see figure \ref{Gen}b).
The number of ways that magnons can hop between the initial and final state - fixed by the boundary condition $\partial$  - is equal to the number possible standard fillings of the generalized Young diagram of figure \ref{Gen}c. For an integral with~$L$ vertices and boundary $\partial$, the OPE leading logarithm is conjectured to be
\beq
\Phi_{\partial} =\frac{(-\log u)^{L}}{L!} \times \texttt{(fishnet stampede)}_{\partial}  \,.
\eeq
A concavity on the left or top edge of the graph correspond to the injection of magnons in the hopping state - i.e. sources - while concavities on the right or bottom edge correspond to the annihilation of a magnon - i.e. sinks.
 \section{{sl(2)} Stampedes and the Octagon}
\label{sect:SL2}
%

The $sl(2)$ Hamiltonian also describes magnons hopping in a spin chain but these magnons are no longer hardcore. They can pile up on top of each other now. We can have $m$ magnons on one site and $n$ magnons on the neighboring site as
\beq
|\underbrace{\overset{\shortstack[l]{$\bullet$\\$\bullet$\\$\bullet$}}{\bullet} }_{m} \underbrace{\overset{\shortstack[l]{$\bullet$\\$\bullet$\\$\bullet$\\$\bullet$\\$\bullet$\\$\bullet$}}{\bullet} }_{n}  \rangle
\eeq
These magnons hop when acted by the Hamiltonian\footnote{Whenever we deal with $\mathcal{N}=4$ SYM correlators and $sl(2)$ stampedes, the coupling $\lambda$ is defined w.r.t. the number of colours $N\to \infty$ and to the Yang-Mills coupling $g_{YM}\to 0$ as
\beq
\lambda = -\frac{g_{YM}^2 N}{16 \pi^2}\,. \la{defLambda}
\eeq
In order to make contact with \cite{AldayBissi}, the coupling there is $a=-4\lambda$, while the coupling used in \cite{Beisert} is $-2\lambda$.  Coherently, our definition of $sl(2)$ Hamiltonian has opposite sign w.r.t. that of \cite{Beisert}. The unusual choice of the minus sign in the definition (\ref{defLambda}) is simply so that the leading logarithms $(\lambda\log(z))^k$ are positive in the light-cone OPE limit $z\to 0$, and at the same time the hopping terms in the hamiltonian -- the key players in this paper -- do not come with annoying minus signs. 
}. They can move in groups. If $d$ magnons hop from a site to the neighbour the Hamiltonian weight is $\lambda/d$; if the magnons do nothing the weight is $-\lambda(h(n)+h(m))$ where 
$
h(n) \equiv \sum_{k=1}^n \frac{1}{k}
$
are the so-called Harmonic numbers. So, for instance, 
\beq
H |\overset{\shortstack[l]{$\bullet$\\$\bullet$}}{\bullet} \overset{\shortstack[l]{$\bullet$}}{\bullet} \rangle= 
\underbrace{\frac{\lambda}{{\color{red}2}} 
|\overset{{\color{gray}\curvearrowleft}}{\overset{\shortstack[l]{{\color{red} $\bullet$}\\{\color{red} $\bullet$}\\$\bullet$\\$\bullet$}}{\bullet}\,  \circ }\rangle
+
\frac{\lambda}{\color{red}1} 
|\overset{{\color{gray}\curvearrowleft}}{\overset{\shortstack[l]{{\color{red} $\bullet$}\\$\bullet$\\$\bullet$}}{\bullet}\,  \bullet} \rangle }_{\texttt{left hopping}}
-\underbrace{\lambda(h(3)+h(2)) |\overset{\shortstack[l]{$\bullet$\\$\bullet$}}{\bullet}  \overset{\shortstack[l]{$\bullet$}}{\bullet} \rangle }_\texttt{no hopping}
+\underbrace{\frac{\lambda}{\color{red}1} |\overset{{\color{gray}\curvearrowright}}{\overset{\shortstack[l]{$\bullet$}}{\bullet} \,\overset{\shortstack[l]{{\color{red} $\bullet$}\\$\bullet$}}{\bullet} }\rangle 
+\frac{\lambda}{\color{red}2} |\overset{{\color{gray}\curvearrowright}}{\overset{\shortstack[l]{}}{\bullet} \,\overset{\shortstack[l]{{\color{red} $\bullet$}\\{\color{red} $\bullet$}\\$\bullet$}}{\bullet}} \rangle
+\frac{\lambda}{\color{red}3} |\overset{{\color{gray}\curvearrowright}}{ \circ\, \overset{\shortstack[l]{{\color{red} $\bullet$}\\{\color{red} $\bullet$}\\{\color{red} $\bullet$}\\$\bullet$}}{\bullet} }\rangle}_\texttt{right hopping} \,. \label{actionH}
\eeq
To define the $sl(2)$ stampede we introduce an incoming state, an outgoing state and a number of time steps. The in state can be thought as a bunch of $J$ particles in a box on the right, occupying a finite portion $K$ of a longer spin chain, and separated by a bridge of length $l$ to another box of length $K$ on the left,
\beq
|\texttt{in}\rangle_\text{open} = |\underbrace{\circ \dots \circ}_{K} \underbrace{\circ \dots \circ}_{l} \texttt{box}_{J,K} \rangle \,,
\eeq
where $\texttt{box}_{J,K} $ is a state of $J$ magnons distributed homogeneously on a chain of length $K$. As illustration a box of size $3$ with $2$ magnons would be given by
\beq
|\texttt{box}_{2,3} \rangle=| \overset{\shortstack[l]{$\bullet$}}{\bullet} \circ \,\circ \rangle+| \circ\overset{\shortstack[l]{$\bullet$}}{\bullet}\,  \circ\rangle +|\circ \circ\,\overset{\shortstack[l]{$\bullet$}}{\bullet}  \rangle+| \bullet \bullet\, \circ \rangle+| \bullet  \circ\, \bullet \rangle+|\circ \bullet \,\bullet \rangle\,.
\eeq
At time $0$ we ``open" the box and let these particle hop to the left. After a few time steps we overlap the resulting state with a state where all particles are now in a box on the right,  
\beq
{}_\text{open}\langle\texttt{out}| = \langle \texttt{box}_{J,K} \underbrace{\circ \dots \circ}_{l} \underbrace{\circ \dots \circ}_{K}   | \,,
\eeq
We assume an open spin-chain with Hamiltonian $H_\text{open}= \sum_{n=1}^{L-1} H_{n,n+1}$ where $H_{n,n+1}$ acts on sites $n$ and~$n+1$ as in (\ref{actionH}) as opposed to a closed spin-chain Hamiltonian where the last particle is connected to the first and leading to an extra term in the Hamiltonian, $ H_\text{closed} = H_\text{open} +H_{L,1}$. With an open Hamiltonian the particles need to move all the way from the right to the left as in a true stampede (they cannot just \textit{go around} the spin chain as in a closed ring). We finally define the $sl(2)$ stampede as 
\beq
{}_\text{open}\langle\texttt{out}| (H_\text{open})^n |\texttt{in}\rangle_\text{open} =\lambda^n (\texttt{open sl(2) stampede})_{K,J,l,n} \,.
\eeq

The \texttt{box} states are $sl(2)$ descendents, obtained by acting on their vacuum state with a bunch of creation operators which spread the magnons homegeneously inside the box. Indeed, one can easily check that
\beq
H |\texttt{box}_{J,K} \rangle =0 \,.
\eeq
up to boundary terms. So at step one the Hamiltonian will only have a non-trivial effect when acting right on the boundary of the box which connects to the bridge. When it does, a few particles from inside the box might hop and leave the box. At a second time step, when we hit the state once more with the Hamiltonian, these particles might move further left in the direction of the left box. Of course, other things will also happen. New particles will get out of the box; some of the particles outside the box will jump inside the box again; etc.
 
Importantly, because of the \textit{bridge} of $l$ empty sites separating the initial to the final box, these particles need to jump at least $l+1$ times before reaching the right. So
\beq
(\texttt{open sl(2) stampede})_{K,J,l,n}  = 
\left\{ \begin{array}{ll} 
O(\lambda^{l+1}) & , \,\, n>l\\
0 &, \,\,  \text{otherwise}
\end{array} \right.
\eeq
Let us outline the general mechanism of this stampedes, leaving a more detailed analysis for the appendix \ref{app:SL2}.
The leading term is simple to compute. Indeed the only way to have a complete stampede of particles from the initial box to the final box is to start from the vector of  $|\texttt{box}_{J,K} \rangle$ with all the $J$ particles on the leftmost site and hop them all to the left at each of the $l+1$ steps:
\begin{figure}[H]
 \begin{center}
\fbox{\includegraphics[trim=1cm 30cm 36cm 0cm,scale=0.5]{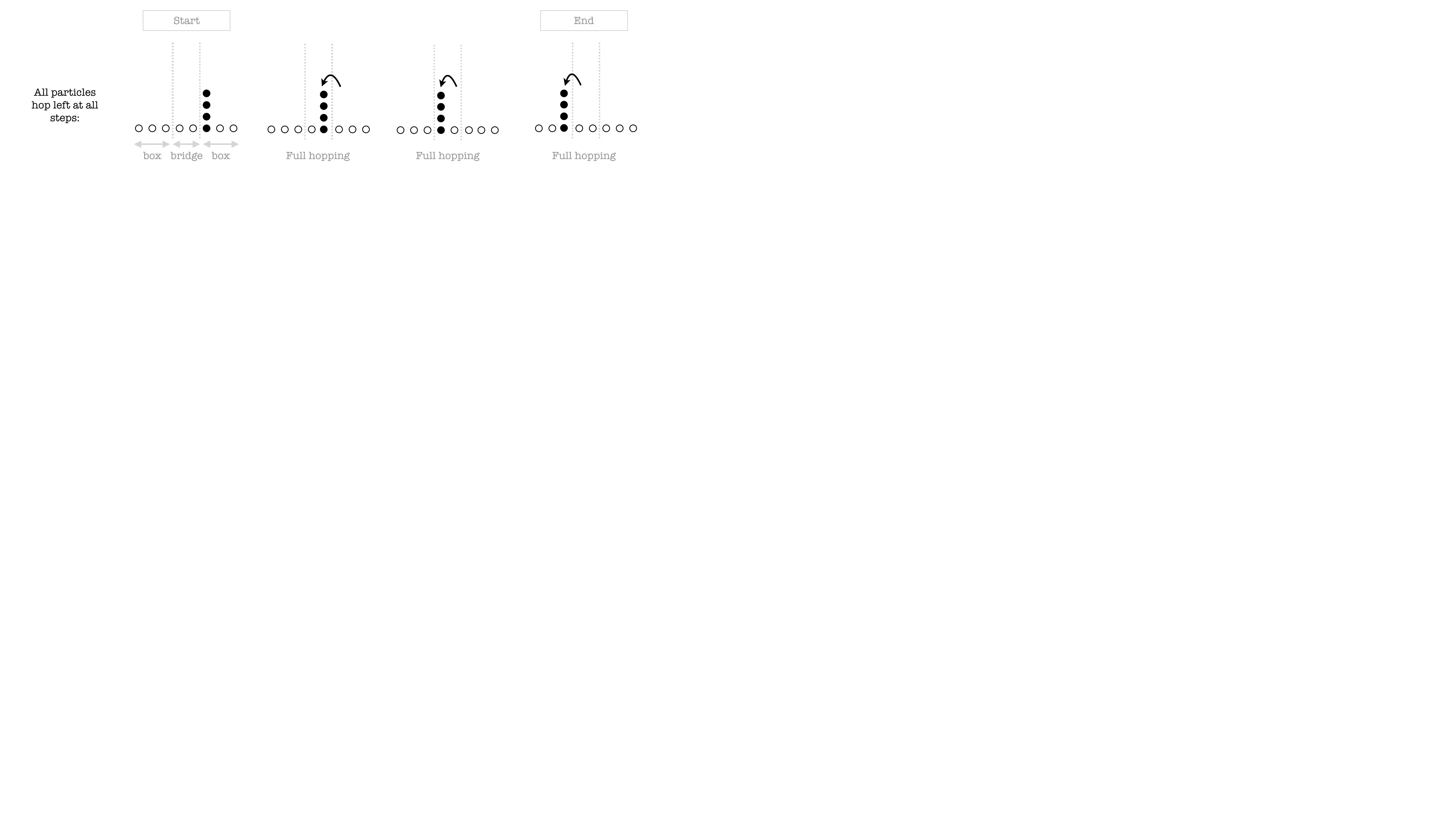}}
\end{center}
 \end{figure}
 \noindent
With the help of the compact notation  \beq
 \label{hop_forward}
 |\underbrace{\circ \dots \circ}_{K+l} \overset{\shortstack[l]{$\bullet$\vspace*{-2mm}\\\hspace*{-1mm} \vdots }}{\bullet} \underbrace{\circ \dots \circ}_{K-1} \rangle\, \equiv \, |\underbrace{\circ \dots \circ}_{K+l} \overset{\shortstack[l]{\vspace*{-0.5mm}\\\hspace*{-1mm} \tiny{J}}}{\bullet} \underbrace{\circ \dots \circ}_{K-1} \rangle\,,
\eeq
the leading order process reads
 \beq
 \label{lead_hop}
H_{\text{open}}^{l+1} |\underbrace{\circ \dots \circ}_{K+l} \overset{\shortstack[l]{\vspace*{-0.5mm}\\\hspace*{-1mm} \tiny{J}}}{\bullet} \underbrace{\circ \dots \circ}_{K-1} \rangle\simeq \frac{\lambda}{J} H_{\text{open}}^{l} |\underbrace{\circ \dots \circ}_{K+l-1} \overset{\shortstack[l]{\vspace*{-0.5mm}\\\hspace*{-1mm} \tiny{J}}}{\bullet} \underbrace{\circ \dots \circ}_{K} \rangle \simeq \cdots \simeq \frac{\lambda^{l+1}}{J^{l+1}} |\underbrace{\circ \dots \circ}_{K-1} \overset{\shortstack[l]{\vspace*{-0.5mm}\\\hspace*{-1mm} \tiny{J}}}{\bullet} \underbrace{\circ \dots \circ}_{K+l} \rangle\,,
\eeq
where the symbol $\simeq$ denotes the projection of the Hamiltonian action on the chosen contributions, where all particles hop left. Thus, the leading term of the stampede is 
\beq
\label{leading_hop}
(\texttt{open sl(2) stampede})_{K,J,l,l+1} = \frac{1}{J^{l+1}} \,.
\eeq
The analysis of sub-leading terms concerns the possible evolutions that bring $|\texttt{in}\rangle_{\text{open}}$ to $|\texttt{out}\rangle_{\text{open}}$ with $n>l+1$ applications of the $sl(2)$ Hamiltonian. This analysis becomes more complicated for bigger $n$, as the number of ways to slow down the stampede grows. Let us consider the next-to-leading case with $n=l+2$ and $K>1$. In order to slow down the leading process by one step there are two possibilities. One possibility is that at one step only a fraction $0\leq r<J$ hop to the left, so that an additional step is needed to hop the remaining $J-r$ particles:
\begin{figure}[H]
 \begin{center}
\fbox{\includegraphics[trim=0cm 24cm 30cm 0cm,scale=0.42]{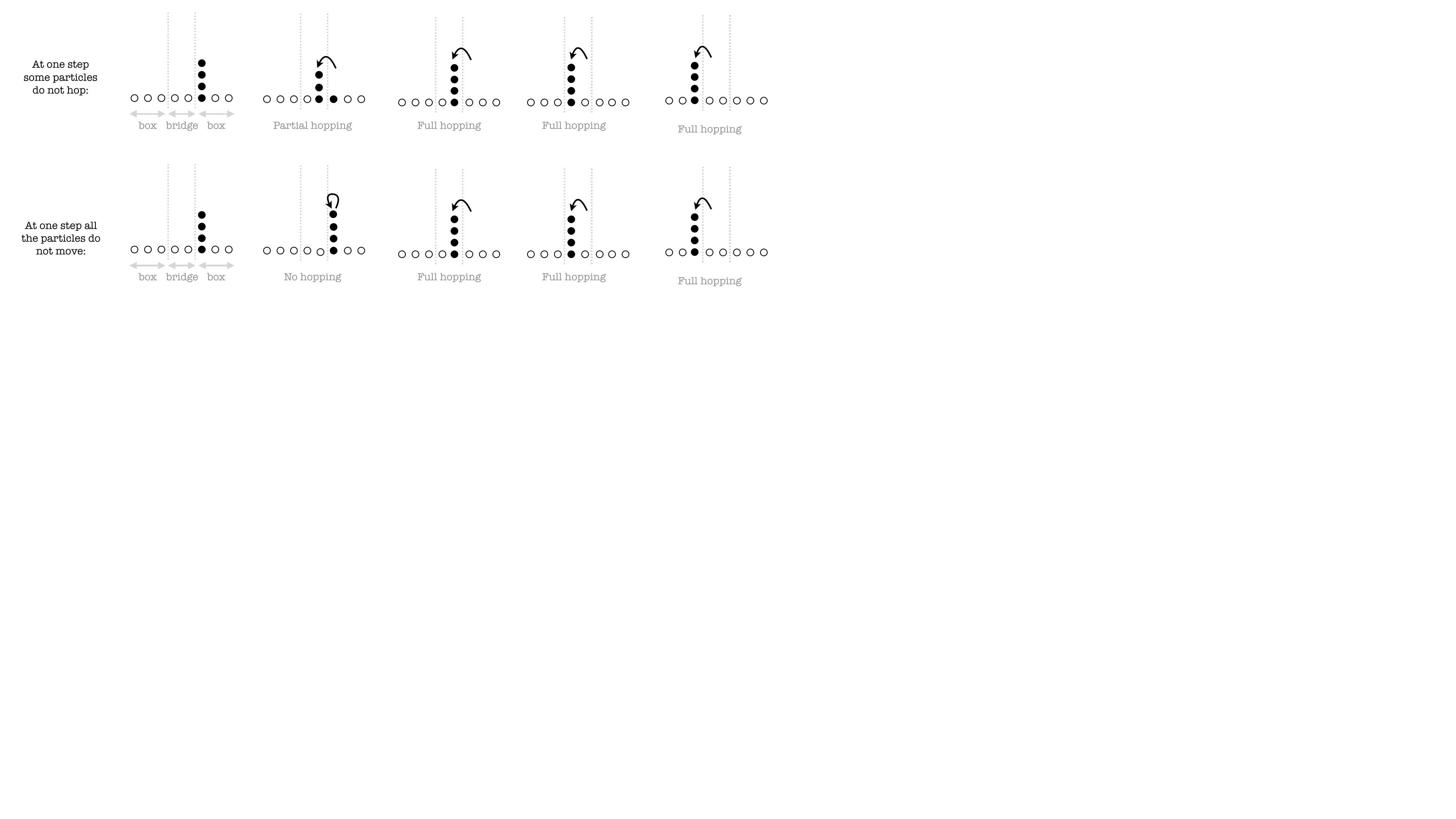}}
\end{center}
 \end{figure}
 \noindent
 The second possibility is if the initial (or final) state has particles distributed in the first (or last) \textit{two} sites of the initial (or final) box, so that one additional step is needed to regroup them in the first (last) site:
 \begin{figure}[H]
 \begin{center}
\fbox{\includegraphics[trim=1cm 30cm 30cm 0cm,scale=0.42]{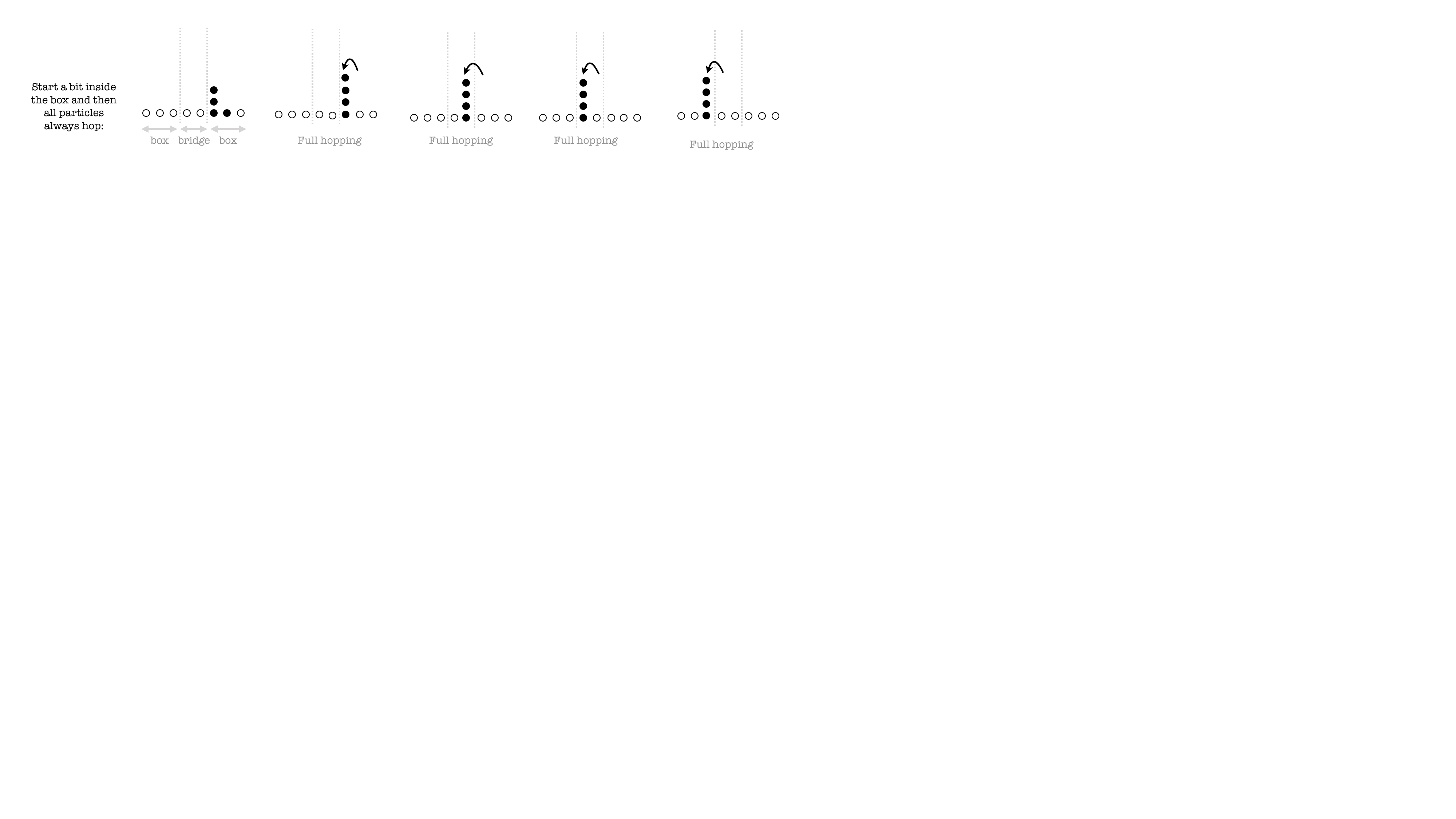}}
\end{center}
 \end{figure}
 \noindent
 Summing over these two contributions, we get the result for the first sub-leading stampede:
\beq
\label{sub_hop}
(\texttt{open sl(2) stampede})_{K,J,l,l+2} = \frac{-2(l+1)}{J^{l+2}} \,,\,\,\,\, K>1\,.
\eeq
Note that this result is independent of $K$ \textit{provided} $K>1$. If $K=1$ boundary effects intervene, as there are no two sites in the box and the contribution of the second kind is not admitted. Moreover, as explained in appendix \ref{app:SL2} since sites $K$ and $K+l+1$ are now at the boundary, the contribution $-2h(J)$ from no-hopping at the first/last step gets halved, and the combined effects preserve the result \eqref{sub_hop}.
This observation ceases to be true at any other sub-leading order $n$, which is affected by boundary terms whenever $2K\leq n-l$. Hence, we concentrate on $K \gg 1$ and define
\beq
\label{open_large_reservoir}
(\texttt{open sl(2) stampede})_{J,l,n}  = (\texttt{open sl(2) stampede})_{\infty, J,l,n}  \,.
\eeq
In practice an efficient way to compute these $(\texttt{open sl(2) stampede})_{\infty, J,l,n}$ is by exploiting some Harmonic polylogarithms technology as explained in appendix \ref{app:HPL}. We summarize here the first few stampedes (the first line is what we just described; in the last line $h_3(J)= \sum_{k=1}^J 1/k^3$) 
\beqa
\begin{aligned}
    &(\texttt{open sl(2) stampede})_{J,l,l+1}=\frac{1}{J^{l+1}} \,,\quad  (\texttt{open sl(2) stampede})_{J,l,l+2}=-\frac{(2l+2)}{J^{l+2}}  \,,\\
     & (\texttt{open sl(2) stampede})_{J,l,l+3}=\frac{(2l+3)(l+2) }{J^{l+3}}  \,,\\
      & (\texttt{open sl(2) stampede})_{J,l,l+4}=- \frac{2 (l+2)(l+3)(2l+5) }{3 J^{l+4}} + 4\delta_{l,0} \Big(\frac{h_3(J)}{J}-\frac{1}{J^4}\Big)  \,.
      \end{aligned}
\eeqa
We come to the main claim of this section:

\textit{The}  (\texttt{open sl(2) stampede})$_{J,l,n}$ \textit{with bridge $l$ govern the leading light-cone limit of the so-called asymptotic octagons $\mathbb{O}_l(z,\bar z)$ introduced by Frank Coronado in \cite{Frank}.} 
%
\begin{figure}[t]
 \begin{center}
{\includegraphics[scale=0.64]{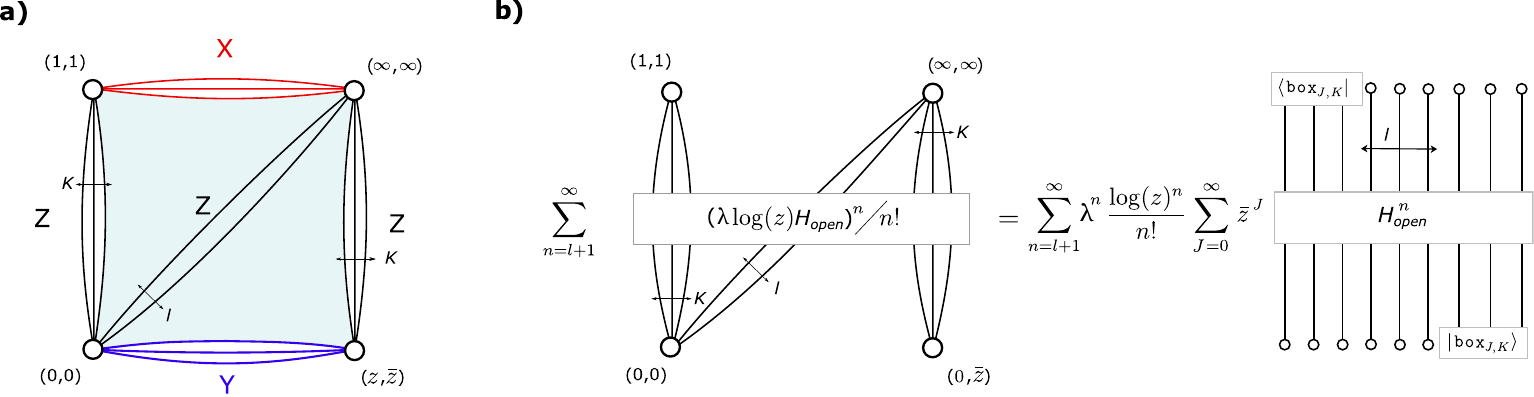}}
\caption{\textbf{a)} Asymptotic octagon with $l$ bridges $\mathbb{O}_{l}(z,\bar{z})$, in light-blue. Solid lines are propagators at the tree level in the corresponding asymptotic correlator. \textbf{b)} Relation between the octagon in light-cone kinematics and the $sl(2)$ stampedes.}
\label{SL2_asymp_fig}
\end{center}
 \end{figure}
 \noindent

The asymptotic octagons are important building blocks of the so-called asymptotic correlation functions \cite{Asymp_4pt,Frank}.
They correspond to the interior of a square frame of $Z$, $X$ and $Y$ fields as indicated in the figure \ref{SL2_asymp_fig}a. Note that in this figure all lines connecting the two bottom operators to the two top operators are made of the same scalar $Z$. This is quite important as states with a single scalar and derivatives in a single light-cone direction form precisely the $sl(2)$ sector discussed so far. When we consider the light-cone OPE between the top and bottom points the $X$ and $Y$ lines in the bottom and top disappear (see figure \ref{SL2_asymp_fig}b). Thus, we end up with a bottom and top effective operators with $Z$ scalars only plus covariant derivatives (since the points are still separated along a null direction). Basically, 
\beq
O^\text{eff}_\text{bottom} \simeq \left( Z^{K+l} e^{\bar z \mathbb{P}} Z^K \right)(0) =\sum_{J=0}^\infty \bar z^J  |\underbrace{\circ \dots \circ}_{K+l} \texttt{box}_{K,J} \rangle\,,
\eeq 
and, similarly
 \beq
O^\text{eff}_\text{top} \simeq  \left( Z^{K+l} e^{\mathbb{K}} Z^K \right) (\infty)= \sum_{J=0}^\infty  \langle  \texttt{box}_{K,J} \underbrace{\circ \dots \circ}_{K+l} | \,,
\eeq 
see also \cite{VW} and \cite{SVW}. We have used that when these derivatives act on any of the last $k$ scalars they define precisely the right \texttt{box} introduced above. 
Finally, to capture the leading logarithm $\log(z)^n$ at order $\lambda^n$ in the light cone limit $z\to 0$ we thus act $n$ times with the one-loop dilatation operator $\mathbb{D}$ on the bottom state before overlapping it with the top state which is precisely what the stampedes do (``loosing" time with any other quantum effect can never produce the maximal $\log(z)$ power).
 In sum, we conclude that in the limit $\log(z) \to \infty$ and $\lambda\to 0$ with their product fixed,  
 \begin{align}
\begin{aligned}
\label{LC_stampede}
\mathbb{O}_{l}(z,\bar z) =  \sum_{n=l+1}^{\infty} \frac{(\lambda \log(z)) ^n}{n!} \sum_{J=1}^{\infty} (\texttt{open sl(2) stampede})_{J,l,n} \,\bar z ^J\,. 
\end{aligned}
\end{align}
This is the main result of this section. 

The \texttt{(open sl(2) stampede)} are very simple objects given combinatorially by counting particles moving from a right box to a left box by action of the one-loop dilation operator. This dilatation operator is integrable in $\mathcal{N}=4$ SYM but that was not really used at any moment. These type of ideas does generalize to any conformal gauge theory with a one-loop dilation operator, be it integrable or not. 

The asymptotic octagon was a particularly nice object since all scalars involved were of a single type $Z$ and thus we could simply use $sl(2)$ stampedes. This is of course not crucial either. Had we considered more general frame with more scalars we would simply need to use more general stampedes where particles have further degrees of freedom -- this is discussed in the appendix \ref{app:general_octagon}. 

Finally, note that we used an open spin chain hamiltonian to define the \texttt{open stampedes} because the side edges in the frames defining the octagon are quite large, with $K\gg 1$ propagators each,  see figure \ref{SL2_asymp_fig}a, which allowed us to decouple the inside and outside of that figure. This is why we could focus only on the blue \textit{interior} of that figure and define the octagon. On the other hand, the number of $X$ and $Y$ lines in that figure does not need to be large. In fact, that number never played any role so far; the results in the single light-cone limit are insensitive to that number of legs. 
In sum, our results as well as this inside/outside decoupling in the single light-cone limits applies to correlators with a frame where the number of side propagators is large but where the bottom and top propagators is arbitrary. 
To study small operator correlators where the number of side propagators is also small we should use a closed Hamiltonian and allow the stampedes to couple dynamically the interior and exterior of the correlator. This is the subject of section \ref{sect:small} below. 

In the next section we will see how the knowledge of the octagon in the double-scaling limit above  (\ref{LC_stampede}) can be used as very powerful boundary data for the octagon bootstrap. 

\section{A Single Light-cone Axiom}

According to \cite{FrankBootstrap} the asymptotic octagon can be cast as
\beq
\label{AsympDeterminant}
\mathbb{O}_l = 1+ \sum_{n=1}^{\infty} 
(z+\bar z-z \bar z)^n \!
\sum_{m=n(n+l)}^{\infty} \lambda^{m}\sum_{\underline{k} 
} \frac{{\color{blue} c^{(l)}_{k_1,\dots,k_n} } }{
\prod\limits_{j=1}^n k_j!(k_j\!-\!1)!}
\underbrace{\begin{vmatrix}
f_{k_{1}}&f_{k_2-1}& \dots & f_{k_n-n+1} \\
f_{k_{1}+1}&f_{k_2}& \dots & f_{k_n-n+2}\\
\vdots&\vdots& \dots & \vdots\\
f_{k_1+n-1}&f_{k_2+n-2}& \dots & f_{k_n}\\
\end{vmatrix}}_{\displaystyle \equiv \mathcal{S}_{\underline{k}}(z,\bar{z})}
\eeq
where the sum over $\underline{k} $ is a sum over the integer partitions $S^{(m,n)}$ of $m$ of length $n$ and where
\beq
\label{ladders}
f_L(z,\bar{z}) =\sum_{j=L}^{2L} \frac{(L-1)!j!}{(j-L)!(2L-j)!} \frac{\Li_{j}(z) - \Li_{j}(\bar{z}) }{z-\bar{z}} (-\log(z\bar{z}))^{2L-j}  \,,
\eeq
are the so-called ladder functions. Fixing the full correlator at all loop orders is thus tantamount to fixing the constants $\color{blue} c^{(l)}_{k_1,\dots,k_n} $. 

The ansatz \eqref{AsympDeterminant} is based on two axioms, first formulated by Frank Coronado in \cite{FrankBootstrap}
\begin{itemize}
\item[\textbf{A.1)}] At loop order $\lambda^{m}$ the octagon is a single valued combination of products of $k<m$ ladder functions $f_{i_1}\cdots f_{i_{k}}$, for $i_1+\dots + i_k=m$. We could call this axiom the \textit{functional basis axiom}. 
\item[\textbf{A.2)}] In the $u$-channel OPE $z\to1, \bar z \to1$ the octagon generates at most single powers of $\log(1-z)$ and $\log(1-\bar z)$. This axiom would naturally be called the \textit{Steimnan relation axiom} since the truncation in single logs is equivalent to stating the vanishing of some double discontinuities, a condition known as the Steimnan relation. Physically it stems from the fact that in this channel we are dominated by the exchange of double-trace operators \cite{StrongOctagon,FrankBootstrap}.
\end{itemize}
Each determinant $\mathcal{S}_{\underline{k}}(z,\bar{z})$ has a leading divergence $\log(z)^m$ for $\underline{k}=(k_1,\dots,k_n) \in S^{(m,n)}$ in the single light-cone limit $z \to 0$ which can be easily read from the light-cone limit of each ladder function  
\beq
f_L(z,\bar{z}) \simeq  {(L-1)!} \frac{\Li_{L}(\bar{z}) }{\bar{z}} \,   (-\log(z))^{L}   \,.
\eeq
All the terms in \eqref{AsympDeterminant} survive the double scaling limit where $\xi = \lambda \log(z)$ is held fixed with $\lambda \to 0^{-}$ and $z\to 0^{+}$. Therefore if we can provide boundary data for the DS limit we can fix completely the constants $c_{\underline{k}}^{(l)}$ and thus the correlator. But this is precisely what the stampedes introduced in the previous section do! The DS limit of \eqref{AsympDeterminant} can be expanded in powers of the coupling as
\begin{align}
\begin{aligned}
 \label{fewloops}
\mathbb{O}_l(\xi;\bar{z}) = 1&+ c_1^{(l)} \log (1-\bar z)\xi + \frac{ c_2^{(l)}\text{Li}_2(\bar z)}{2}\xi^2- \frac{c_3^{(l)} \text{Li}_3(\bar z)}{6} \xi^3+\\&+  \left(c_{1,3}^{(l)}  \left(2 \Li_1(\bar{z}) \Li_3(\bar{z}) - \Li_2(\bar z)^2\right)+   \frac{c_4^{(l)} \text{Li}_4(\bar z)}{24} \right)  \xi^4 + \\&+  \left(\frac{ c_{1,4}^{(l)}}{72}  \left(\Li_2(\bar{z}) \Li_3(\bar{z}) + 3 \log(1-\bar{z}) \Li_4(\bar z)\right)+   \frac{c_5^{(l)} \text{Li}_5(\bar z)}{24} \right)  \xi^5 +\dots \,,
\end{aligned}
\end{align}
and matched against the stampede prediction \eqref{LC_stampede} 
\beqa
\label{LC_stampede}
\mathbb{O}_l(\xi;\bar{z}) =\sum_{m=l+1}^{\infty} \frac{\xi^m}{m!} \sum_{J=0}^{\infty} (\texttt{open sl(2) stampede})_{J,l,m} \,\bar z ^J\,,
\eeqa
so that by simply evaluating the sum over the stampedes computed in the previous section we can  easily read off the $c$'s. We can list the first few $c_{i}^{(l)}$ obtained this way in a nice table:
\begin{center}
\begin{tabular}{c|cccccccccccccc}
 \backslashbox{$l$}{$i$} & $1$ & $2$ & $3$ & $4$& $\{1,3\}$ & $5$ & $\{1,4\}$ & $6$ & $\{1,5\}$ & $\{2,4\}$ \\\hline
$0$ & $1$ & $-2$ & $6$ & $-20$ & $1$  & $70$  & $-6$  & $-252$ & $28$ & $18/5$  \\
$1$ & $0$ & $1$ & $-4$ & $15$ & $0$  & $-56$ &  $0$ & $210$ & $0$& $1$ \\
$2$ & $0$ & $0$ & $1$ & $-6$ & $0$  & 28 &   $0$ & $-120$ & $0$&$0$   \\
\end{tabular}
\end{center} 
In other words, the last axiom allowing us to completely bootstrap all $\mathbb{O}_l$ is to require that:
\begin{itemize}
\item[\textbf{A.3)}] In the \textit{single} light-cone double-scaling limit ($\lambda \log(z)$ fixed with~$z,\lambda\to 0$) the octagon~$\mathbb{O}_l$ is given by the stampedes. 
\end{itemize}
The first line in this table perfectly agrees with the constants fixed by Coronado using a different axiom which however is only valid for zero bridges $l=0$:
\begin{itemize}
\item[\textbf{A.3 FC)}] The logarithm octagon $\mathbb{O}_0$ in the \textit{double} light-cone limit $z\to 0$ and $\bar z\to \infty$ generates at most quadratic terms in the logarithm of these cross-ratios. 
\end{itemize}
This axiom is much simpler and much more powerful to implement than the stampede axiom. 

Still, it has its own drawbacks which the stampedes axiom does not. First, there is not derivation of this axiom or even a strong argument for it. Instead, the quadratic truncation in the double light-cone limit was observed from collecting lots of integrability data and then interpreted as arising from the absence of recoil \cite{Alday:2010zy,FrankBootstrap,AldayBissi,StrongOctagon,BK0,BGV,BGHV} in the double null limit  where correlators make contact with Wilson loops \cite{Alday:2010zy}. Secondly, and more pragmatically, it only holds for $l=0$. The quadratic truncation in the double light-cone limit is not observed for $l>0$ which is where the stampedes are thus most useful. 
re
Would be formidable if we could find an optimal formulation of the third axiom which would be as powerful as Coronado's one is but valid for non-zero bridges as well. Hopefully the stampedes can provide some helpful intuition. 

In that vain, one possible direction would be to explore a very nice double-scaling limit which sits somewhere in between the two limits in \textbf{A.3} and \textbf{A.3 FC} and which was introduced by Belitsky and Korchemsky in \cite{BK}. In this limit we take a double light-cone limit as in  \textbf{A.3 FC} but we further tune the cross-ratios and the coupling as in  \textbf{A.3}, in this case to keep 
\beq
s^2=\lambda \log(z) \log(\bar z) \la{sVar}
\eeq
fixed. They found that  in this limit the octagons simplify to the beautiful expression 
\beq
 \frac{\mathbb{O}_l }{ \mathbb{O}_0} \simeq  \det_{i,j\le l} I_{|i-j|}(2s)\qquad \texttt{\underline{double} LCDS limit} \, . \label{detI}
\eeq
This \textit{double} scaling limit is a particular case of the \textit{single} scaling limit in \textbf{A.3} and can be reached from the later by further taking the leading term as $\bar z$ becomes large so this nice determinant of Bessel functions is a particular case of the stampedes.\footnote{\footnotesize{
The double-scaling limit selects the leading behaviour in $\bar z\to \infty$ from the light-cone octagon. In fact, in this limit the stampedes are dominated by the processes where particles never move to the right. That is, in~(\ref{actionH}) we can drop the \texttt{right} hopping terms and still land precisely on \eqref{detIC}. We call these simpler stampedes generated by that simplified Hamiltonian as \texttt{chiral sl(2) stampedes}. For example, let's look at the expansion until five loops in $\xi=\log(z)\lambda$ for the light-cone octagon with $l=1$ bridge
\beq
\mathbb{O}_1\simeq \mathrm H_{2}(\bar z)\frac{\xi^2}{2!}  - 4  \mathrm H_{3}(\bar z) \frac{\xi^3}{3!} +
(13 \mathrm H_{4}(\bar z) - 2 \mathrm H_{2, 2}(\bar z)) \frac{\xi^4}{4!} + 
  (\!-40 \mathrm H_{5}(\bar z) \!+ 4 \mathrm H_{2, 3}(\bar z) + \!12\mathrm H_{3, 2}(\bar z)) \frac{\xi^5}{5!}\,.
\eeq
The result for open chiral stampedes differs already at three loops
\begin{align}
\begin{aligned}
\mathbb{O}^{\text{chiral}}_1&\simeq  \mathrm H_{2}(\bar z) \frac{\xi^2}{2!} + 4  \mathrm H_{2, 1}(\bar z) \frac{\xi^3}{3!} + 
  (7 \mathrm H_{2, 2}(\bar z) + 2 \mathrm H_{3, 1},(\bar z) + 24 \mathrm H_{2, 1, 1}(\bar z)) \frac{\xi^4}{4!} +\\&\qquad\,\,\,\,\,+
  (52 \mathrm H_{2, 1, 2} (\bar z) + 60 \mathrm H_{2, 2, 1}(\bar z) + 
    24 \mathrm H_{3, 1, 1}(\bar z) + 192 \mathrm H_{2, 1, 1, 1}(\bar z))\frac{\xi^5}{5!}\,.
\end{aligned}
\end{align}
Nevertheless, taking the null-square double-scaling limit, one nicely checks that indeed $\mathbb{O}_1/\mathbb{O}^{\text{chiral}}_1 \to 1$.}
\small
} One could wonder if we could fix the constants $c_i^{(l)}$ by matching with such gorgeous  expression in this double light-cone limit. This is not possible. Starting at four loops this only fixes some linear combinations of the constants. On the other hand we could take the single light-cone limit used in \textbf{A.3}, take $\bar z$ large but retain not only the leading term but the full $1/\log(\bar z)$ expansion. As explained in appendix \ref{app:toda} this defines a $\tau$-function which allows us to generalize (\ref{detI}) to a more general solution of the Toda hierarchy which reads
\beq
\label{tauDet}
 \frac{\mathbb{O}_l }{\mathbb{O}_0} \simeq  \det_{i,j\le l} \left[
 \Big(\frac{\partial}{\partial L}\Big)^{i-1}  \Big(\frac{\partial}{\partial \bar L}\Big)^{j-1} \tau(L,\bar L)
 \right] \qquad \texttt{\underline{single} LCDS limit} \,,
 \eeq
 where 
 \beq
L \equiv \lambda \log(z) \,\, , \qquad \bar{L} \equiv \lambda^0 \log(1/\bar{z}) \la{singleDS}\,,
\eeq
and
\beq
\tau = I_0 \big(2 \sqrt{L\bar L}\big) + \left(\frac{L}{\bar L}\right) \frac{\pi ^2}{6 } \left(I_2\big(2 \sqrt{L\bar L}\big) -2 s \,I_1\big(2 \sqrt{L\bar L}\big) \right) + \dots
 \eeq
We computed the dots in this expression up to order $(L/\bar L)^6$ but we were not able to fully resum this expression into a nice closed form. The expression $\tau(L,\bar L)$ and \eqref{tauDet} cannot be used to completely bootstrap the all-loop octagon, while it fixes it until $8$ loops, where the information of single {DS} double light-cone limit is still enough to disentangle all coefficients~$c_i^{(l)}$ entering the ansatz \eqref{AsympDeterminant}.

\section{Small Correlators}
\label{sect:small}
So far we studied large operators in $\mathcal{N}=4$ SYM but clearly the very same stampedes apply to small operators as well, provided we use a \textit{closed} stampede -- where the particles can hop all around the sites of a closed chain -- rather than two decoupled \textit{open} stampedes describing the two decoupled octagons present for large operators. Let us illustrate this with a simple example which should make the general case obvious. Consider a four point function of these operators:
\beq
\mathcal{O}_1= \Tr(ZZZY) \,, \qquad 
\mathcal{O}_2= \Tr(Z\bar{Y}) \,, \qquad
\mathcal{O}_3= \Tr(\bar{Z}\bar{X}) \,, \qquad
\mathcal{O}_4= \Tr(\bar{Z}\bar{Z}\bar{Z}\bar{X}) \,. 
\eeq
When we take $z\to 0$ we take the leading null OPE so we can replace\footnote{At this point we are not keeping track of the overall normalization of any operators. We fix this normalization below by fixing the tree level contribution $f_0$ with respect to the loop contributions $f_{n>0}$.} 
\beq
\mathcal{O}_1 \mathcal{O}_2 \to \Tr(Z^3 e^{\bar z \mathbb{P}_-} Z) = \sum_{J=0}^\infty \bar{z}^J \Big(
\underbrace{
|\circ\circ\circ \underbrace{\overset{\shortstack[l]{$\bullet$\\$\bullet$\\$\bullet$}}{\bullet} }_{J}  \rangle+
|\circ\circ \underbrace{\overset{\shortstack[l]{$\bullet$\\$\bullet$\\$\bullet$}}{\bullet} }_{J}  \circ\rangle+
|\circ \underbrace{\overset{\shortstack[l]{$\bullet$\\$\bullet$\\$\bullet$}}{\bullet} }_{J} \circ\circ \rangle+| \underbrace{\overset{\shortstack[l]{$\bullet$\\$\bullet$\\$\bullet$}}{\bullet} }_{J} \circ\circ\circ  \rangle
}_{|\texttt{in}\>_J }\Big)\,,
\eeq
and similarly
\beq
\mathcal{O}_3 \mathcal{O}_4 \to \Tr(\bar Z^3 e^{\mathbb{K}_-} \bar Z)=  \sum_{J=0}^\infty  \Big(
\underbrace{\< \circ\circ\circ \underbrace{\overset{\shortstack[l]{$\bullet$\\$\bullet$}}{\bullet} }_{J}  |+
\<\circ\circ \underbrace{\overset{\shortstack[l]{$\bullet$\\$\bullet$}}{\bullet} }_{J}  \circ|+
\<\circ \underbrace{\overset{\shortstack[l]{$\bullet$\\$\bullet$}}{\bullet} }_{J} \circ\circ |+
\< \underbrace{\overset{\shortstack[l]{$\bullet$\\$\bullet$}}{\bullet} }_{J} \circ\circ\circ  | 
}_{{}_J\<\texttt{out}| }\Big)\,.
\eeq
All we did is to contract the $Y$ and $X$ fields -- which are spectators when we take the null limit $z\to 0$ -- and then write the effective bottom and top states in a manifestly cyclic symmetric way. Then we simply have 
\beq
\<\mathcal{O}_1 \mathcal{O}_2 \mathcal{O}_3  \mathcal{O}_4\> \simeq  \frac{1}{x_{12}^2 x_{34}^2x_{13}^2 x_{24}^2(x_{23}^2)^2}\times \sum_{n=0}^\infty \frac{(\lambda \log(z))^n}{n!}\,\overbrace{\sum_{J=0}^\infty \bar z^J\,\underbrace{ {}_J\<\texttt{out}| (H_\texttt{closed})^n |\texttt{in}\>_J }_{(\texttt{closed sl(2) stampede})_{J,n}}}^{\displaystyle  f_n(\bar z)} \la{prediction}
\eeq
It is then a simple matter of combinatorics to act with the closed Hamiltonian, hop the derivaties and compute any $f_n$. For instance, we find for the first ones 
\beqa
f_0&=&3+\frac{ 1}{1-\bar z}\,, \notag \\ 
f_1&= &\frac{2 \bar z\,}{\bar z-1} \textrm{H}_{1}(\bar z) \,, \notag\\
f_2&= &-\frac{6 \bar z\, \textrm{H}_{2}(\bar z)+4 (\bar z+1) \textrm{H}_{1,1}(\bar z)}{\bar z-1} \,, \notag\\
f_3&= &\frac{20 \bar z\, \textrm{H}_{3}(\bar z)+6
   (\bar z+5) \textrm{H}_{1,2}(\bar z)+4 (4 \bar z+5) \textrm{H}_{2,1}(\bar z)+8 (\bar z+5) \textrm{H}_{1,1,1}(\bar z)}{\bar z-1}\,, \\
f_4&=& \frac{-72\bar z \,\textrm{H}_{4}(\bar z)+16 (\bar z-11)
   \textrm{H}_{1,3}(\bar z)+2\bar z \left(7 \textrm{H}_{2,2}(\bar z)-2 \textrm{H}_{3,1}(\bar z)+6 \textrm{H}_{1,1,2}(\bar z)\right)}{\bar z-1}+\notag\\&+ &\frac{2\bar z \left(6 \textrm{H}_{1,2,1}(\bar z)-4
   \textrm{H}_{2,1,1}(\bar z)+8 \textrm{H}_{1,1,1,1}(\bar z)\right)-5 (23 \textrm{H}_{2,2}(\bar z)+18 \textrm{H}_{3,1}(\bar z))}{\bar z-1}+\notag\\&+&\frac{10(30 \textrm{H}_{1,1,2}(\bar z)+30 \textrm{H}_{1,1,2}(\bar z)+30
   \textrm{H}_{1,2,1}(\bar z)+28 \textrm{H}_{2,1,1}(\bar z)+40 \textrm{H}_{1,1,1,1}(\bar z))}{\bar z-1}\,.\notag
\eeqa
where the functions $\textrm{H}_{a,b,c,\dots}(\bar z)$ are the standard harmonic polylogarityhms (HPLs) \cite{HPL} defined in the appendix \ref{app:HPL}. Here we point out that the combination of HPLs that appear in $f_{n\leq 4}$ can be actually rewritten in terms of classical polylogs $\Li_k(\bar z)$ only, whereas starting from $f_5$ and for higher loops this property does not hold. This is an interesting difference w.r.t. the case of the open stampede, where only classical polylogs appear at any loop order, in agreement with the fact that octagon functions are expressed as determinants of ladder integrals \eqref{AsympDeterminant},\eqref{ladders}.
\begin{figure}[t]
\begin{center}
\includegraphics[scale=0.5]{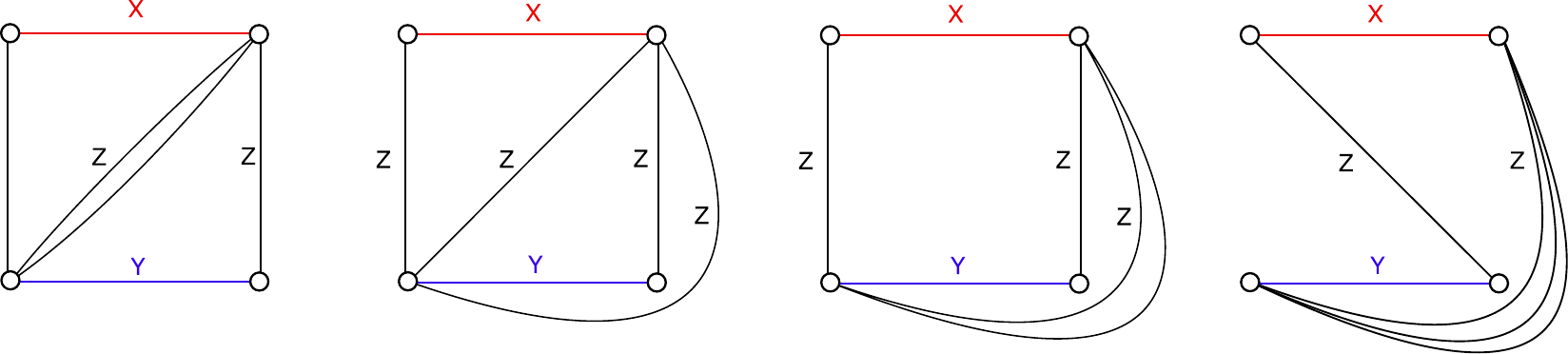}
\end{center}
\caption{Planar skeleton diagrams at tree level of the correlator $\<\mathcal{O}_1 \mathcal{O}_2 \mathcal{O}_3  \mathcal{O}_4\>$.}
\label{skeletons}
\end{figure}
We can now take the large $\bar z$ limit so that the four points are approaching the corners of a null square. In particular, if we keep only the leading term
\begin{align}
\begin{aligned}
&f_0\simeq 3\,,\quad   f_1 \simeq  -2 \log (\bar z) \,,\quad  f_2\simeq   \log ^2(\bar z) \,, \quad
f_3 \simeq -  \log^3(\bar z) \,, \quad 
f_4 \simeq \frac{49}{12} \log^4(\bar z) \,,
\end{aligned}
\end{align}
we can plug these expressions into (\ref{prediction}) to obtain the prediction 
\beq
\<\mathcal{O}_1 \dots \mathcal{O}_4\> \simeq \frac{1}{x_{12}^2 x_{34}^2x_{13}^2 x_{24}^2(x_{23}^2)^2} \times \left(3- 2 s^2+ \frac{s^4}{2} -\frac{s^6}{6} + \frac{49}{288} s^8 +\dots \right) \,,
\eeq
where $s$ is defined in (\ref{sVar}). We note that this nicely resums into 
\beq 
\label{octagons_products}
\<\mathcal{O}_1 \dots \mathcal{O}_4\> \simeq \frac{1}{x_{12}^2 x_{34}^2x_{13}^2 x_{24}^2(x_{23}^2)^2} \times \left(2\, \mathbb{O}_2(2s) \mathbb{O}_0(2s)+ \mathbb{O}_1(2s)  \mathbb{O}_1(2s)\right)\,,
\eeq
where $\mathbb{O}_n(2s)/\mathbb{O}_0(2s)$ are the Bessel function determinants (\ref{detI}) showing up in the previous section. This fact can be nicely understood. At tree-level the correlator is given by four skeleton graphs depicted in figure \ref{skeletons}. The first three are distinguished planar graphs, as they differ by the position of the two bridges $1/x_{23}^2$, but evaluate to the same expression, while the last is responsible for the second summand in $f_0$ and it is suppressed when $\bar z \to \infty$. 
The correlator at loop order is obtained decorating the skeleton diagrams with interactions. In the double light-cone limit $x_{12}^2,x_{24}^2,x_{34}^2,x_{13}^2 \to 0$, the internal and the external region of the square frame of each skeleton decouple into the product of two simplet objects - the octagons. (Note that the fourth diagram which was suppressed when $\bar z \to \infty$ is dropped at all loops.) Due to the decoupling, each region could be computed via open stampedes with large reservoirs and a number of bridges $l$ equal to the number of propagators $1/x_{23}^2$ appearing in that region. This precisely produces a null octagon $\mathbb{O}_l(2s)$, explaining \eqref{octagons_products} with all the various octagon bilinears with its precise numerical prefactors.

\begin{figure}[t]
\begin{center}
\includegraphics[scale=0.36]{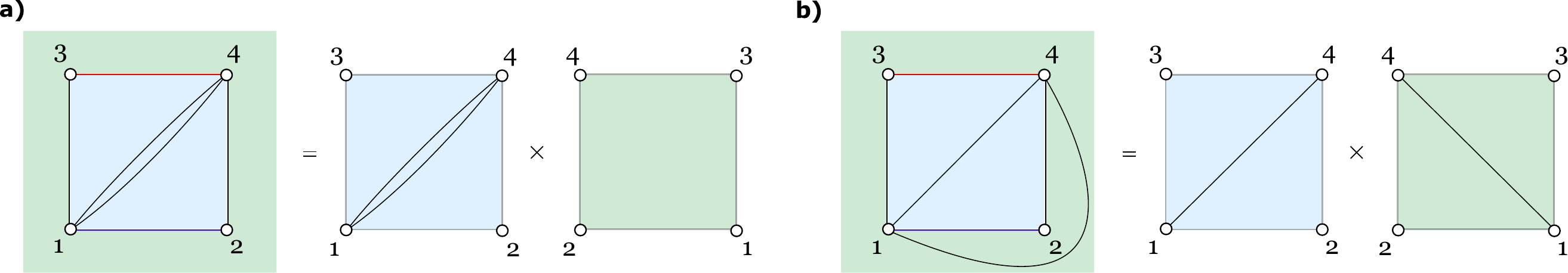}
\end{center}
\caption{Factorization of the contributions corresponding to a certain skeleton diagram. The interior and exterior of the square frame are identified with two octagons that decouple in the null-square limit. \textit{a)} and $\textit{b)}$ correspond to the first two diagrams of figure \ref{skeletons}.}
\label{skelet_fact}
\end{figure}

Following this same logic, it is straightforward to make a prediction for the simpler correlator of the lightest protected operators, also known as $\textbf{20'}$. In this case the relevant skeleton has no diagonal bridges and is thus  simply given by the square of the empty octagon in the double scaling limit, 
\beq
\<\Tr\left(XZ\right )  \Tr\left(\bar XZ\right )  \Tr\left(Y\bar Z\right )  \Tr\left(\bar Y \bar Z\right )   \> \simeq \frac{e^{-2s^2}}{x_{12}^2 x_{13}^2 x_{24}^2 x_{34}^2}\,,\,\,\,\,\, \mathbb{O}_0(2s) =e^{-s^2}\,.
\eeq

\begin{figure}[t]
\begin{center}
\includegraphics[trim=0cm 3cm 0cm 0cm, scale=0.35]{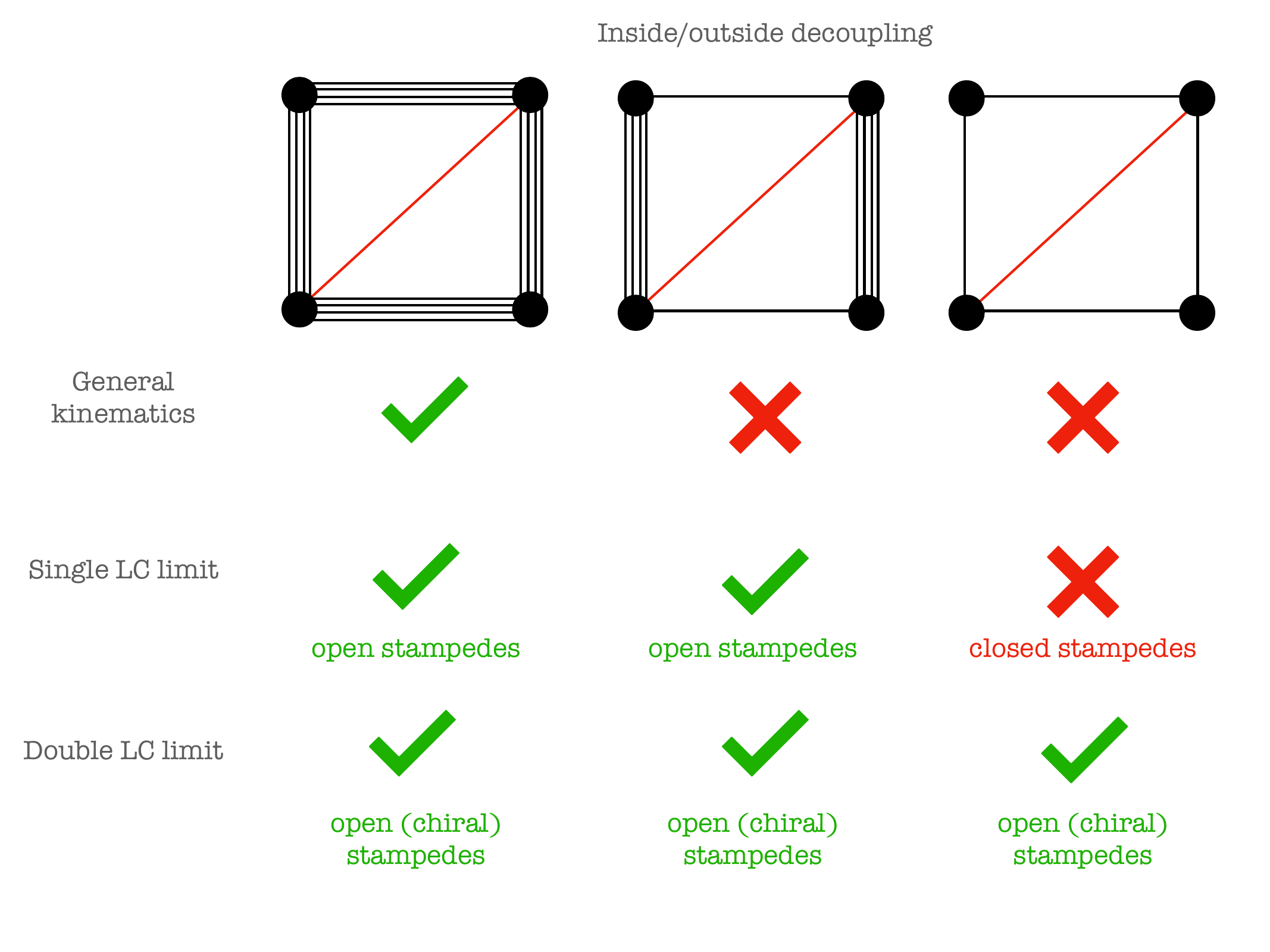}
\end{center}
\caption{Summary of decoupling conditions. In the first line we show square skeleton diagrams for different thickness of the edges. We distinguish the condition of factorization into octagons by the symbol $\color{green} \checkmark \color{black}$, while not factorizable cases are marked by $ \color{red}  \times \color{black}$. As long as we take at least a single double scaling null limit (i.e. for the last two lines of the figure) we can predict the outcome through the stampedes even when there is no decoupling as summarized in the figure.}
\label{cases_null}
\end{figure}

In sum, a four-point correlator of $\frac{1}{2}$-BPS operators factorizes into octagon functions at all loops whenever it gets contributions only from skeleton graphs with a square frame of propagators whose edges are subject to a decoupling condition. There are two such conditions. The first, standing at the core of the work \cite{Frank}, is when a number $K\gg1$ of propagators is stretched along that edge. The second, is when an edge lies on the light-cone, and we take (an appropriate) double-scaling limit of the correlator. Any mix of the two conditions along the square frame is possible, as showed in figure \ref{cases_null}.
\section{Discussion and Future Problems}
Several simple dynamical processes whereby particles move through the repeated action of a local Hamiltonian from a fixed \texttt{in} state to another fixed \texttt{out} state were considered in this paper. We called them \texttt{stampedes}. In large $N$ quantum field theories, such local Hamiltonians often appear in the planar limit and by considering the action of these Hamiltonians -- and nothing else -- we isolate so-called \textit{leading log contributions} in these theories. When the logs are large and the coupling is small  we can sometimes define double scaling limits fully dominated by the \texttt{stampedes} which we can analyse at all loop orders by simply counting paths for the particles to move from the \texttt{in} to the \texttt{out} state. A very similar leading log analysis in the null polygonal Wilson loop context was considered in \cite{SVW}.

The simplest quantum field theory we studied here was the fishnet field theory \cite{Fishnet}. This is quite a simple theory with two complex scalars and a single quartic coupling $g \Tr(XZ \bar X \bar Z)$. In this theory correlators with disk topology (i.e. single-trace correlators) are given by a single Feynman diagram which behaves as
\beq
\includegraphics[scale=0.3]{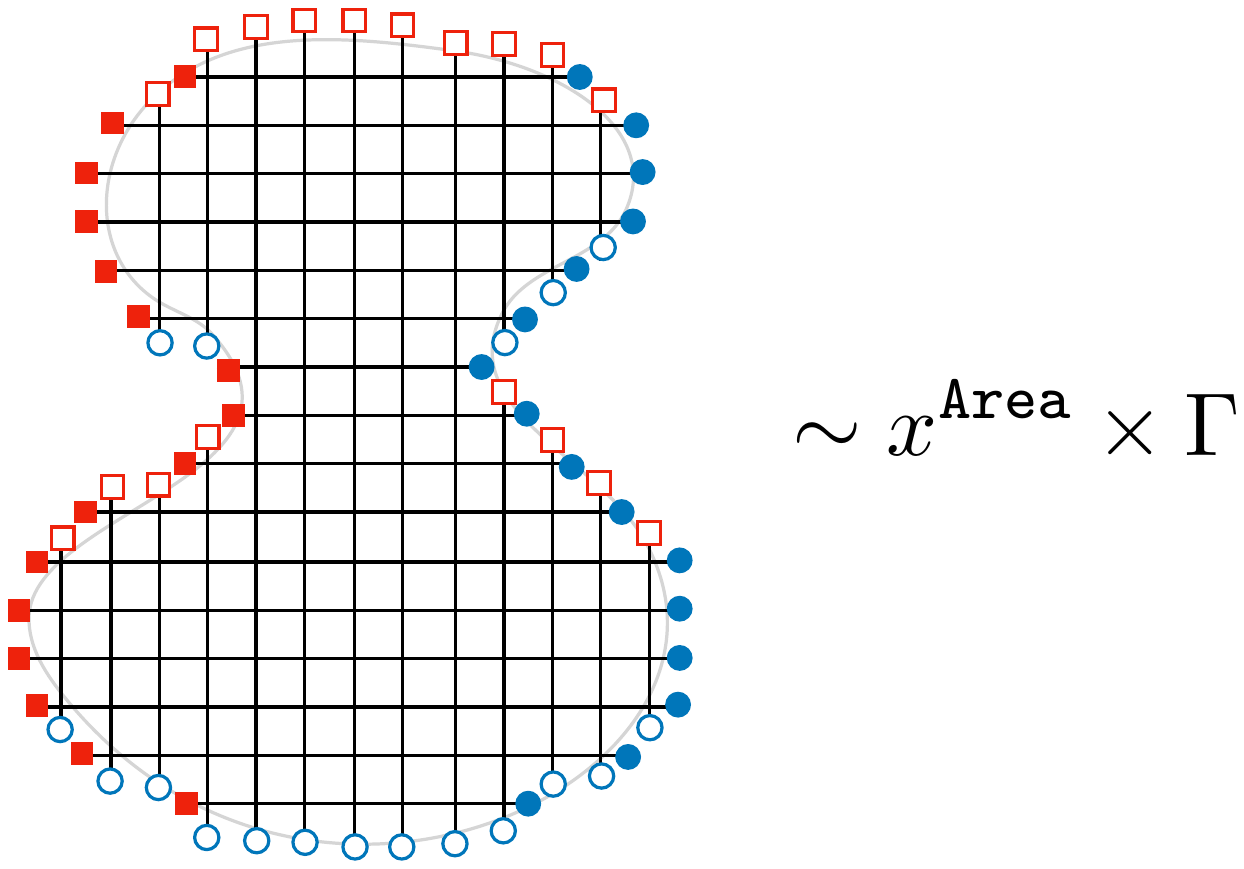}
\vspace{-1.5cm} \nn
\eeq
where $x= g \log(z\bar z)$ is the double scaling variable mentioned above, \texttt{Area} is the number of loops, i.e. four point vertices in the disk, and where $\Gamma$ is equal to the number of \texttt{chiral stampedes} connecting red (circles) and blue (square) portions of the shore as explained in sections \ref{sect:intro} and \ref{sect:fishnet}. Counting such paths is a simple problem which can be nicely mapped to that of counting (generalized) Young tableaux as explained above. It would be interesting to find out what happens in the continuum limit of huge disks. In \cite{BassoDixonAndFriends} it was emphasized that the boundary conditions matter in the large area limit of the fishnet, so we expect this limit to be quite rich. Will we get some kind of coherent stampede a fluid flow of sorts?

Even for small disks, the counting of paths we explained here provide interesting predictions for future fishnet evaluation of these Feynman diagrams. Can we generalize the rectangle results of Basso and Dixon in \cite{BassoDixon} to general disk topologies? The technology developed in \cite{Derkachov:2019tzo,Derkachov:2020zvv,Derkachov:2021rrf, Olivucci:2021cfy} should be powerful enough for this task. For concave disks we expect some sort of Jacobi-Trudi like generalizion of their nice determinant results where some indices of columns or rows are appropriately shifted. 

Next we considered \texttt{sl(2) stampedes} where particles still move from an in to an out state but now the particles can move back and forth along the way, they can move in packs, and can also share the same site along the way. We can think of these stampedes as starting with a boson gas in some incoming box, opening it at some initial time and collecting the particles in another box at some later time. We saw that these stampedes govern the leading single light-cone limit of  four point correlation functions in large N conformal gauge theories. We considered $\mathcal{N}=4$ SYM as an important application. We saw that these stampedes can be used as boundary data and compared with the bootstrap program of Frank Coronado \cite{Frank} to fix an infinite family of correlators of large R-charge to all loop orders.

\begin{figure}[t]
\begin{center}
{\includegraphics[scale=0.9]{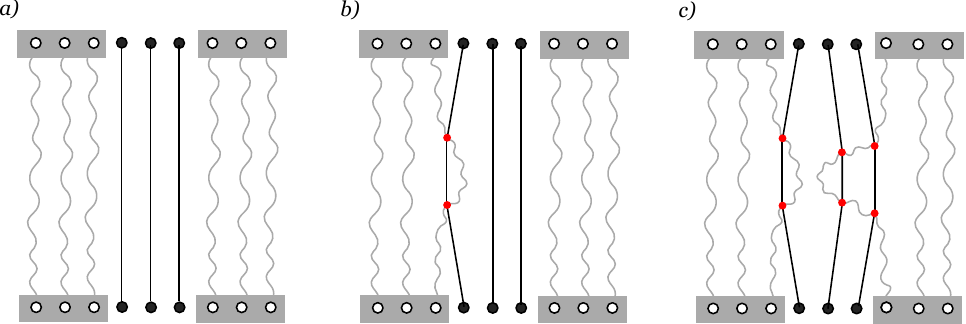}}
\caption{Left/right hopping of $l=3$ hard magnons/bridges in presence of reservoirs of holes/gluons. \textbf{a)} $m=0$ steps, no interaction, order $s^0$.  \textbf{b)} $m=2$ steps, emitted and absorbed magnon, order $s^2\simeq \lambda$. \textbf{c)} $m=6$ steps. The order in which the vertices occur in the time-direction (up-down) matters and counts different paths. The number of such paths is $ (\texttt{bridge stampedes})_{l,m}$ discussed in the text.}
\label{hop_glue}
\end{center}
\end{figure}

We then considered the double light-cone limit which is dominated by large spin in the field theory. In the stampede language this limit is dominated by processes with many particles moving from an \texttt{in} to an \texttt{out} box. The stampedes in this case allow to make contact with beautiful closed formulae previously known \cite{BK}. In the double scaling limit where (\ref{sVar}) is fixed, for instance, we find a nice determinant of Bessel functions 
\beq
 \label{detIC}
 \frac{\mathbb{O}_l }{ \mathbb{O}_0} \simeq  \det_{i,j\le l} I_{|i-j|}(2s)\qquad \texttt{\underline{double} LCDS limit} \, .
\eeq
governing the double light-cone limit of the octagons with bridge length $l$. The result (\ref{detIC}) seems to hint at yet other stampedes which we call \texttt{bridge stampedes} which are yet to be understood. Indeed, it is a known combinatoric result \cite{Grabiner} that 
%
\beq
\det_{1\leq i,j,\leq l} I_{i-j}(2 s) = \sum^\infty_{m=0} \frac{s^m}{m!} (\texttt{bridge stampedes})_{l,m}\,,
\eeq
where  $(\texttt{bridge stampedes})_{l,m}$ is the number of random walks of $m$ steps $y_k \mapsto y_k \pm 1$, in the interior of the Weyl chamber of the lattice $\mathbb{Z}^l$ (i.e. for hardcore excitations), starting and ending at the same point $(l-1,l-2,\dots,0)$
\footnote{As an example $ \sum\frac{s^n}{n!} \int dt \overbrace{(e^{it}+e^{-it}) \dots (e^{it}+e^{-it})}^\texttt{n times}=\int_0^{2\pi} \frac{dt}{2\pi} e^{2s \cos(t)}= I_0(2s)$. The first sum is designed to be a generating function for the origin-to-origin paths for a single particle. Indeed, 
to get a non-zero integral we must cancel any power of $e^{\pm i t}$ which means that if we expand out all terms we keep those where we select as many $e^{it}$ as $e^{it}$. If we think of each factor in the integrand as a step and each such choice of $e^{\pm i t}$ as a choice of stepping left or right at that step then it is clear that by selecting the same number of left and right steps we are counting all paths which come back to the origin. Similarly, $I_m(2s)=\sum\frac{s^n}{n!} \int dt e^{i m t} {(e^{it}+e^{-it})^n} $. To get a non-zero result we must still cancel all exponentials but now we have the extra $e^{im t}$, so we conclude that we are counting the number of paths that start at a position $m \in \mathbb{Z}$ and end at the origin after a bunch of steps. This is all trivial for a single particle. What is quite remarkable is that for many hardcore particles there still exists such a nice generating function, the determinant of Bessel functions \cite{Grabiner}. The starting idea is to think of these many hardcore particles in the line as a single particle moving in a higher dimensional Weyl chamber. The number of paths in an $l$ dimensional Weyl chamber can be related to alternating sums of the numbers of \textit{unconstrained} paths in $\mathbb{Z}^l$ by some beautiful reflection tricks by Gessel and Zeilberger \cite{GZ}; basically, the carefully chosen signs in the alternating sums ensure that all paths cancel in pairs except for those in the Weyl chamber which are the ones we are after. Now, the unconstrained paths are products of Bessel functions by the same kind of logic of the single particle in a line with which we started this footnote. The sign-alternating sums of such products then nicely combine into determinants~\cite{Grabiner}. 
}
\beq
| \dots \circ \circ\, \underset{l}{\underbrace{\bullet\dots \bullet}} \circ \circ \dots  \rangle\,.
\eeq
 We may think of this hopping as the planar interactions of the bridges with the null frame. The latter can only emit and absorb gluons, therefore we can regard it as a reservoir of gluons (empty spaces) standing on the left and right of the the bridges (i.e. the magnons), as illustrated in figure \ref{hop_glue}. We see that the leading null-square behaviour of a $n$-point BPS correlator in $\mathcal{N}=4$ SYM differs, for $l>0$ bridges, from a null polygonal Wilson loop by a generating function for the number of certain random walks, related to the way the bridges can interact with the null frame. Of course, this is a (very hand-waving) interpretation. Can we make this rigorous and propose an alternative combinatorial derivation of the double scaling limit result (\ref{detIC})? Can we also shed light over the more general result (\ref{tauDet}) which we found here? We know from \cite{Determinant,BK0,BK} that the light-cone Octagon may be regarded as a probability in a determinantal process with a Bessel kernel extended with a Fermi-Dirac distribution. It would be nice to fully understand the stochastic picture behind it, beyond the zero-temperature case corresponding to the null double-scaling limit. In particular, do the corrections in (\ref{tauEq}) also have a random walk interpretation? 
(The fact that the same process can be described by an $sl(2)$ process involving derivatives or an hardcore procewss involving the scalars of the bridges reminds of the nice background independence description of Basso and Dixon fishnet diagrams \cite{BassoDixon}: their results can both be understood from a flux tube expansion or from a BMN expansion \cite{BassoDixon}.)
We refer to a few intriguing papers \cite{OkounkovReshetikhin,LeDouss,Johnasson,Baik,MNS}, and to the relevant references therein, that may shed light to a deeper stochastic interpretation of the octagon functions.

Stampedes also predict the leading null limit of small correlation functions. The space of functions describing small correlators is much larger so it is not obvious how to use the stampede predictions as boundary data in an efficient analytic bootstrap program for small operators. Maybe in some simplifying kinematical restriction (where we put points in a single line say) we can make progress in this direction. It would be very interesting to explore this further. It was also recently understood that all the scalar correlators in $\mathcal{N}=4$ SYM exhibit an enhanced $10d$ symmetry \cite{FrankSimon}. This should manifest somehow in the stampede language. 

We could imagine using the stampedes to learn about more virgin territory. We have no hexagonalization formalism in ABJM, for instance, but we could easily define the stampedes there and produce infinite set of predictions for null limits of correlators therein. Can we get some hints about an integrable description of correlation functions in this three dimensional theory? 

Finally, very little is known about higher point correlation functions of five or more points (not even in $\mathcal{N}=4$ SYM or in ABJM). In multiple null limits, some progress has been made in bootstrapping these correlators in \cite{BGV, BGHV, MiguelEtAl}. Would be nice to reproduce the double scaling limit of these results -- with logs of cross-ratios exploding and the coupling vanishing with appropriate products held fixed -- from stampedes and also to generalize them to more general correlators with internal bridges. Is there an analogue of the beautiful Bessel determinant~(\ref{detIC}) for the five and six point functions? That could be quite useful boundary data for ongoing studies of so-called \texttt{decagons} and \texttt{dodecagons} \cite{decagons} which intend to generalize Frank Coronado's \texttt{octagons}. In principle, all these polygons come from gluing together \texttt{hexagons}~\cite{hexagons, Hexagonalization} but in practice only for the octagon do we know how to do this efficiently \cite{Frank}. Getting nice stampede results for the other cases could hint at important hidden structures in the hexagon formalism more generally which could significantly improve this state of affairs.

 \section*{Acknowledgements}
We would like to thank Carlos Bercini, Frank Coronado, Vasco Goncalves, Alexandre Homrich, Vladimir Kazakov, Zohar Komargodski, Henrique Malavazzi and Amit Sever for illuminating discussions. We would also like to thank Eitan Bachmat for pointing out relevant references and interesting random walks applications related to sections \ref{sect:intro} and \ref{sect:fishnet}. We also thank Frank Coronado for sharing with us data about octagons.
Research at the Perimeter Institute is supported in part by the Government of Canada 
through NSERC and by the Province of Ontario through MRI. This work was additionally 
supported by a grant from the Simons Foundation (Simons Collaboration on the Nonperturbative Bootstrap \#488661) and ICTP-SAIFR FAPESP grant 2016/01343-7 and FAPESP grant 2017/03303-1. The work of P.V. was partially supported by the Sao Paulo Research Foundation (FAPESP) under Grant No. 2019/24277-8.  E.O. would like to thank FAPESP grant 2019/24277-8 for funding the visit to ICTP-SAIFR (FAPESP grant: 2016/01343-7) from October to December 2021 where part of this work was done.
\newpage
\appendix 
\section{Stampedes and Young tableaux}
\label{app:Catalan}
The problem of counting the number of ways that the $m$ hard magnons of section \ref{sect:intro} can move each of $n$ steps in one direction, is a famous combinatorial problem \cite{Gessel}. Its solution can be mapped to the counting of standard Young tableaux (SYT). Let us consider a rectangular Young diagram of size $m \times n$. The hopping process requires $mn$ steps, labelled by integers $i=1,2,\dots,m n$. Each of the $m$ rows of the Young diagram is associated to a magnon, and its $n$ boxes shall be filled with the ordered list of steps $\{i_1,\dots, i_n\}$ in which that magnon hops left. For the sake of clearer correspondence between the graph and the tableaux, we fix the order to be decresing along rows/columns. The fact of magnons being hard and hopping only left defines a bijection between walks and rectangular SYTs of size $(m\times n)$. The problem of counting SYTs of a certain shape - defined by the length of its rows $\{\lambda_1,\dots,\lambda_m\}$ - is given by the formula \cite{Frobenius}
\beq
\label{Hook_length}
(\#\text{SYT})_{\lambda_1,\dots,\lambda_m} = \frac{(\lambda_1+\dots +\lambda_m)!}{\prod_{(i,j)}h_{\underline{\lambda}}(i,j)}\,,
\eeq 
where $(i,j)=(\text{row},\text{column})$ of a box in the diagram, and $h_{\lambda_1,\dots,\lambda}(i,j)$ is the number of boxes in the ``hook" defined by $\{(i,j'\geq j)\} \cup \{(i'\geq i,j)\}$. For the rectangular diagrams of interest we have $\lambda_i=\lambda_j=n$, and the length of a hook is easily determined as $h_{n,\dots n}(i,j)= (n+m-i-j+1)$, thus formula \eqref{Hook_length} becomes
\beq
\label{Catalan_higher}
(\#\text{SYT})_{n,n,\dots,n} = \frac{(n m)!}{\prod_{i=1}^m\prod_{j=1}^n (m+n-i-j+1)}\,.
\eeq 
These numbers go under the name of Catalan numbers in $m$ dimensions. For $m=2$ they are the standard Catalan numbers $C_n$, and count the ways a particle can walk from $(x_1,x_2)=(1,0)$ to $(n+1,n)$ in the interior of the Weyl chamber $x_1> x_2$ of $\mathbb{Z}^2$ in $2n$ steps of type $x_k \to x_k+1$. For $m>1$ they count the walks from $(x_1,x_2,\dots,x_m)=(m-1,m-2,\dots,0)$ to $(m+n-1,m+n-2,\dots,n)$ in $mn$ steps of type $x_k \to x_k+1$, inside the higher-dimensional Weyl chamber on $\mathbb{Z}^m$.
The r.h.s. of \eqref{Catalan_higher} can be written via the Barnes' $G$-function \cite{BassoDixonAndFriends}
\beq
(\#\text{SYT})_{n,n,\dots,n} = (mn)! \frac{G(m+1)G(n+1)}{G(m+n+1)}\,,
\eeq
that is the form they appear in the leading OPE order of Basso-Dixon integrals as in (4.21) of \cite{BassoDixonAndFriends}. 

\begin{figure}[t]
\begin{center}
\setlength{\fboxsep}{2pt}
\setlength{\fboxrule}{1pt}
{\includegraphics[scale=0.4]{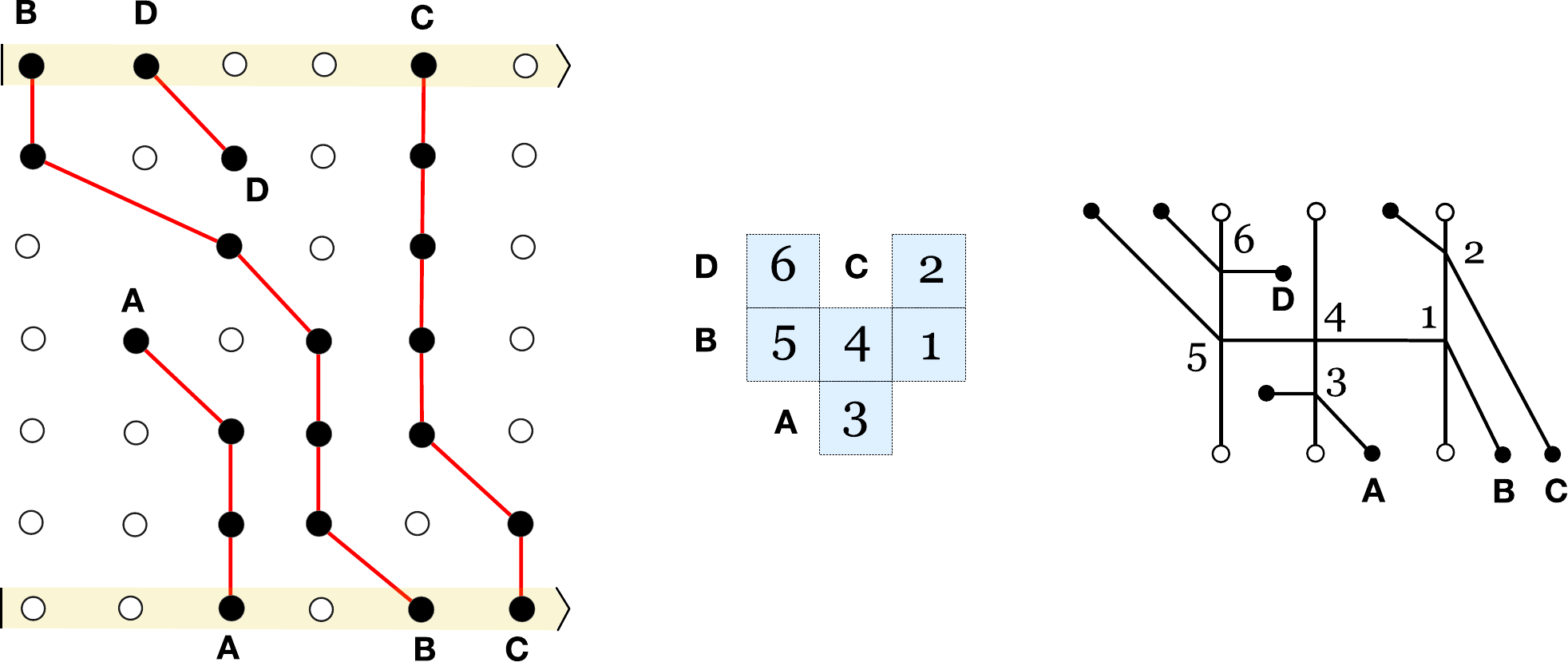}}
\end{center}
\caption{A possible way of hopping of $m=4$ hard magnons - labelled as $A,B,C,D$ - with annihilation of $C$ and injection of $D$, corresponding to a generalized Young diagram, i.e. to a concave fishnet on the disk.}
\label{counting_gen}
\end{figure}

The counting explained so far can be extended to any shape of a fishnet integral on the disk, namely any portion of infinite square-lattice that can be cut out drawing a circle that defines a boundary $\partial$. In this case the conjectured leading $\log$ OPE coefficient is given by the counting of the fillings of a generalized Young diagram obtained replacing vertices with boxes, such that along each row/column the integers are decreasing as for the SYT prescription. We provide a simple example in figure \ref{counting_gen} corresponding to the Fishnet CFT correlator
\beq
\< \Tr\left[ X^2 (x_1) Z(x_2) X(x_3) Z(x_4) \bar X(x_5)  Z(x_6) \bar X^2(x_7) \bar Z(x_8) X (x_9) \bar Z (x_{10}) \bar X (x_{11})  \bar Z (x_{12})  \right] \>\,,
\eeq
and notice that along the stampede magnons are injected or annihilated at a precise stage, in correspondence to the concavities of the diagram. When all fields $X,Z$ and $\bar X,\bar Z$ are sent respectively to two space-time points, the integral of figure \ref{counting_gen} shall diverge as \beq
\Phi_{\partial}\simeq \frac{(-\log u)^6}{6!} \times 9\,,
\eeq 
where $u$ is a cross-ratio of the twelve-point correlator such that $u\to 0$ in the limit.
\section{Harmonic Polylogarithms}
\label{app:HPL}
In section \ref{sect:SL2} we observed that $\sum_{J=1}^{\infty} x^J (\texttt{open sl(2) stampede})_{J,l,n}$ can be expressed as a linear combination of harmonic polylogarithms $\textrm{H}_{\underline{a}}(x) = \textrm{H}_{a_1,\dots,a_m}(x)$ \cite{HPL}. The Harmonic Polylogarithms \cite{HPL} are analytic functions in a neighbourhood of $x=0$, thus we define their Taylor series around this point. For the special case $m=0$ the function is trivial, while for $m=1$ they reduce to the standard polylogarithms
\beq
\textrm{H}_{{0}}(x) = 1\,,\,\,\,\,\, \textrm{H}_{a}(x) = \text{Li}_a(x) = \sum_{j=1}^{\infty} \frac{x^j}{j^a}\,,\,\,\,\,\, |x|<1\,.
\eeq
In order to define the general $\textrm{H}$, we need to introduce the Euler-Zagier harmonic sums
\beq
Z_a(j) = \sum_{k=1}^{j-1} \frac{1}{k^a}\,,
\eeq
and their recursive generalization to any number $m\geq1$ of integer parameters $a_i>0$
\beq
Z_{\underline{a}} (j) =Z_{a_1\dots a_m} (j) = \sum_{k=1}^{j-1} \frac{Z_{a_2,\dots,a_m}(k)}{k^{a_1}}\,.
\eeq
The $\textrm{H}$ are defined around $x=0$ by the series expansion
\beq 
\textrm{H}_{a_1\dots a_m}(x)=\sum_{j=1}^{\infty}  \frac{Z_{a_2,\dots,a_m}(j)}{j^{a_1}} \,x^j\,,\,\,\,\,\, |x|<1\,.
\eeq
and by analytic continuation otherwise. 
An important property of HPLs is their asymptotic behaviour at $x\gg 1$
\beq
\label{HPLasympt}
\textrm{H}_{a_1\dots a_m}(x) = (-1)^m \frac{(\log x )^{a_1+\dots+a_m}}{(a_1+\dots+a_m)!}\,.
\eeq 
This property is frequently used in section \ref{sect:small}, where the limit $\bar z\to \infty$ of the functions $\textrm{H}_{\underline{a}}(\bar{z})$ captures the null-square limit of the of the octagon functions.
 \section{sl(2) stampede: detailed computations}
 \label{app:SL2}
Let us consider the sl(2) stampede initial state $|\texttt{in}\rangle_{\text{open}}$, defined by the $J$ particles all on the site $K+l+1$. The action of the $sl(2)$ Hamiltonian gets non-zero contributions by the local interaction on sites $(K+l,K+l+1)$ and $(K+l+1,K+l+2)$, respectively given by
 \beq
  \label{hop_12}
 -2 \lambda h(J)|\underbrace{\circ \dots \circ}_{K+l} \overset{\shortstack[l]{\vspace*{-0.5mm}\\\hspace*{-1mm} \tiny{J}}}{\bullet} \underbrace{\circ \dots \circ}_{K-1} \rangle+\lambda \sum_{k=1}^J\frac{1}{k}|\underbrace{\circ \dots \circ}_{K+l-1}  \overset{\shortstack[l]{\vspace*{-0.5mm}\\\hspace*{-1mm} \tiny{k}}}{\bullet} \overset{\shortstack[l]{\vspace*{-0.5mm}\\\hspace*{-1mm} \tiny{J-k}}}{\bullet} \underbrace{\circ \dots \circ}_{K-1} \rangle \,,
\eeq
and
 \beq
   \label{hop_01}
 -2 \lambda h(J)|\underbrace{\circ \dots \circ}_{K+l} \overset{\shortstack[l]{\vspace*{-0.5mm}\\\hspace*{-1mm} \tiny{J}}}{\bullet} \underbrace{\circ \dots \circ}_{K-1} \rangle+\lambda \sum_{k=1}^J\frac{1}{k}|\underbrace{\circ \dots \circ}_{K+l}  \overset{\shortstack[l]{\vspace*{-0.5mm}\\\hspace*{-1mm} \tiny{J-k}}}{\bullet} \overset{\shortstack[l]{\vspace*{-0.5mm}\\\hspace*{-1mm} \tiny{k}}}{\bullet} \underbrace{\circ \dots \circ}_{K-2} \rangle \,.
\eeq
If $l=0$ this concludes the analysis, as only one vector has non-zero overlap with $_{\text{open}}\langle \texttt{out} | = \cdots +\langle \underbrace{\circ \dots \circ}_{K-1} \overset{\shortstack[l]{\vspace*{-0.5mm}\\\hspace*{-1mm} \tiny{J}}}{\bullet} \underbrace{\circ \dots \circ}_{K+l} |+\cdots $, and any other vector in \eqref{hop_12} and \eqref{hop_01} does not. The former is the vector where all particles have moved one site towards the right
\beq
\label{leading_hop}
\frac{\lambda}{J}|\underbrace{\circ \dots \circ}_{K+l-1} \overset{\shortstack[l]{\vspace*{-0.5mm}\\\hspace*{-1mm} \tiny{J}}}{\bullet} \underbrace{\circ \dots \circ}_{K} \rangle\,.
\eeq
Moreover, a new application of $H_{\text{open}}$ would raise the power of $\lambda$ giving sub-leading terms, i.e. terms where the hopping of particles towards the left of the chain is slower. If $l=1$, the first non-zero overlap with $_{\text{open}}\langle \texttt{out} |$ is obtained after $l+1=2$ applications of the Hamiltonian to  $| \texttt{in}\rangle_{\text{open}}$. The action of $H_{\text{open}}$ on \eqref{leading_hop} delivers it
\beq
\label{leading_hop}
\frac{\lambda^2}{J^2}|\underbrace{\circ \dots \circ}_{K-1} \overset{\shortstack[l]{\vspace*{-0.5mm}\\\hspace*{-1mm} \tiny{J}}}{\bullet} \underbrace{\circ \dots \circ}_{K+1} \rangle\,.
\eeq
Conversely, the action of $H_{\text{open}}$ on any other vector of \eqref{hop_12}, \eqref{hop_01} overlaps to zero, since there is no way to move $J-k\geq 1$ derivatives by two sites (towards left) via a nearest-neighbour interaction.
If $l>1$ the scheme can be iterated, as we recognize that \eqref{leading_hop} is proportional to the state $|\texttt{in}\rangle_{\text{open}}$ with the replacements $K\to K+1$ and $l\to l-2$. Thus, we conclude that the leading logarithm for $l$ bridges is given by
\beq
 {}_{\text{open}}\langle \texttt{out}| H_{\text{open}}^{l+1} |\texttt{in}\rangle_{\text{open}} ={}_{\text{open}}\langle \texttt{out}|\underbrace{\circ \dots \circ}_{K-1} \overset{\shortstack[l]{\vspace*{-0.5mm}\\\hspace*{-1mm} \tiny{J}}}{\bullet} \underbrace{\circ \dots \circ}_{K+l} \rangle \frac{\lambda^{l+1}}{J^{l+1}} = \frac{\lambda^{l+1}}{J^{l+1}}\,.
\eeq
The analysis of sub-leading terms involves all the ways of getting from $|\texttt{in}\rangle_{\text{open}}$ to $|\texttt{out}\rangle_{\text{open}}$ with $n>l+1$ applications of the $sl(2)$ Hamiltonian. The analysis gets more complicated with growing $n$, as this increases the range of possibilities to slow down the hopping of particles towards the left. Let us consider $n=l+2$ and $K=1$. The only way to achieve non-zero overlaps with $|\texttt{out}\rangle_{\text{open}}$ is to slow down the stampede at one step, either splitting the hopping of $J$ particles from a site $k$ to a site $k-1$ into two passages, or letting the particles not move at one step. In this way we start and end with the very same configurations as in the leading order, but the coefficient produced is different:
\beq
 \label{sub_leading}
 \frac{\lambda^{l+2}}{J^l}\left[ \sum_{r=1}^{J-1} \frac{(l+1)}{(J-r)r} - (2l+2) \frac{h(J)}{J} \right]\, =\, -\frac{2(l+1)}{J^{l+2}}  \lambda^{l+2}\,.
 \eeq
 If we consider $K>1$, at order $n=l+2$ we must take into account that some non-zero overlap comes also from those in and out vectors where a number of particles are not on the side adjacent to a bridge, but on the next one. This adds contributions of the type
 \beq
  \label{boundary_sub1}
H_{\text{open}}^{l+2}\sum_{r=1}^{J}|\underbrace{\circ \dots \circ}_{K+l}  \overset{\shortstack[l]{\vspace*{-0.5mm}\\\hspace*{-1mm} \tiny{J-r}}}{\bullet}\overset{\shortstack[l]{\vspace*{-0.5mm}\\\hspace*{-1mm} \tiny{r}}}{\bullet}  \underbrace{\circ \dots \circ}_{K-2} \rangle\,\simeq\, H_{\text{open}}^{l+1}\sum_{r=1}^{J}\frac{\lambda}{r} |\underbrace{\circ \dots \circ}_{K+l}  \overset{\shortstack[l]{\vspace*{-0.5mm}\\\hspace*{-1mm} \tiny{J}}}{\bullet}  \circ \underbrace{\circ \dots \circ}_{K-2} \rangle \simeq h(J) \frac{\lambda^{l+2}}{{J^{L+1}}} |\underbrace{\circ \dots \circ}_{K-1}  \overset{\shortstack[l]{\vspace*{-0.5mm}\\\hspace*{-1mm} \tiny{J}}}{\bullet}  \underbrace{\circ \dots \circ}_{K+l} \rangle\,,
 \eeq
 and
 \beq
 \label{boundary_sub2}
H_{\text{open}}^{l+2}|\underbrace{\circ \dots \circ}_{K+l}  \overset{\shortstack[l]{\vspace*{-0.5mm}\\\hspace*{-1mm} \tiny{J}}}{\bullet} \underbrace{\circ \dots \circ}_{K-1} \rangle\,\simeq\,  \frac{\lambda^{l+1}}{J^{l+1}}H_{\text{open}} |\underbrace{\circ \dots \circ}_{K-1}  \overset{\shortstack[l]{\vspace*{-0.5mm}\\\hspace*{-1mm} \tiny{J}}}{\bullet}  \underbrace{\circ \dots \circ}_{K+l} \rangle \simeq \frac{\lambda^{l+2}}{{J^{L+1}}} \sum_{r=1}^J \frac{1}{r}  |\underbrace{\circ \dots \circ}_{K-2}  \overset{\shortstack[l]{\vspace*{-0.5mm}\\\hspace*{-1mm} \tiny{J-r}}}{\bullet} \overset{\shortstack[l]{\vspace*{-0.5mm}\\\hspace*{-1mm} \tiny{r}}}{\bullet}  \underbrace{\circ \dots \circ}_{K+l} \rangle\,,
 \eeq
 where by $\simeq$ we mean that we are picking only the relevant terms for the overlap that have not already been taken into account for $K=1$. Moreover, the contributions $h(J)$ due to the particles standing still at the first step shall be doubled, as it can come both also the action of the hamiltionian on sites $(K+l,K+l+1)$ and $(K+l+1,K+l+2)$, while for $K=1$ there is no site $K+l+2$ to act on. The same doubling happens for particles standing still at the last step. Therefore, summing $-2h(J)\lambda^{l+2}/J^{l+1}$ to contribution \eqref{boundary_sub1} $+$ \eqref{boundary_sub2}$= 2h(J)\lambda^{l+2}/J^{l+1}$ and to \eqref{sub_leading} does not change the result.
 
The example we discussed for $n=l+2$ shows that the result is not affected by the size $K$ of the box, even though the possible hoppings are sensible to it. Nevertheless starting at $n=l+3$ this ceased to be true, and the results for $K=1$ and $K=2$ differ. In particular, if the size is big enough, $2K>r$ at the sub-leading order $n=l+r$, the boundary effects disappear and the result is independent of $K$. The particles need to hop across the $l$ empty sites but they cannot reach the boundary of the box and bounce back in a final state with non-zero overlap with $|\texttt{out}\rangle_{\text{open}}$. We can therefore concentrate on very large $K$, and define
\beq
(\texttt{open sl(2) stampede})_{J,l,n}  = (\texttt{open sl(2) stampede})_{\infty, J,l,n}  \,.
\eeq

\section{Light-cone OPE vs hopping: $\mathbf{20'}$ operators}
\label{app:20p}
In this section we provide a check of stampede techniques for a four-point correlator of the lightest protected operators, also known as $\mathbf{20}'$ operators $\mathcal{O}_i = \Tr\left(y_i \cdot \Phi \right)(x_i)$. 
With an appropriate choice of polarization vectors $y_i$ we fix the correlator to be
\beq
\label{20p_4pt}
\langle\Tr\left( X Z \right)(x_1)\Tr\left(\bar X Z \right)(x_2)\Tr\left(\bar Z Y \right)(x_3)\Tr\left(\bar Z \bar Y \right)(x_4) \rangle = \frac{\mathcal{G}(z, \bar z)}{ x_{12}^2  x_{34}^2 x_{13}^2  x_{24}^2 }\,,
\eeq
where $z\bar z =(x_{12}^2x_{34}^2)/(x_{13}^2x_{24}^2)$, $(1-z)(1-\bar z) =(x_{14}^2x_{23}^2)/(x_{13}^2x_{24}^2)$, 
and we consider the leading behviour in the light-cone limit $x_{12}^2,x_{34}^2\to 0$. The latter is captured by the OPE over all the primary operators $O_J$ of minimal twist $\tau(\lambda)=\Delta(\lambda)-J = 2 +\gamma_{1}(J) \lambda+O(\lambda^2)$, and any even spin $J\in \mathbb{N}$, that leads to the $s$-channel conformal blocks expansion \cite{DolanOsb, AldayBissi}
\beq
\label{LC_OPE_light}
\mathcal{G}(z,\bar z) \simeq  \sum_{J\,\text{even}} 2 \frac{\Gamma(1+J)^2}{\Gamma(1+2J)} (z\bar z)^{\frac{\gamma_{1}(J)}{2} \lambda} \bar z^J {}_2F_1\left(1+J,1+J,2+2J ;\bar z \right)\,.
\eeq
The spectrum of anomalous dimensions is simple at one-loop and reads
\beq
\gamma_1(J) = -8h(J)\,, \label{gammaS}
\eeq
and it is the only quantum correction relevant for the leading $\log$ in the light-cone regime. We can test the expansion over \emph{closed} spin-chain states of $sl(2)$ magnons, against \eqref{LC_OPE_light}; for this we write
\begin{align}
\begin{aligned}
&\Tr\left(X  Z \right)(1,1)\Tr\left(\bar X  Z \right)(\infty ,\infty) \simeq \Tr(\bar Z e^{\mathbb K_-}\bar Z )(\infty,\infty) =\sum_{J=0}^{\infty} \langle \overset{J}{\bullet} \circ |\,,\\ & \Tr\left(Y \bar Z \right)(0,0)\Tr\left(\bar Y \bar Z \right)(0,\bar z) \simeq \Tr(Z e^{\bar z \mathbb P_-}Z )(0,0) =\sum_{J=0}^{\infty}\bar z ^J |\circ \overset{J}{\bullet} \rangle \,,
\end{aligned}
\end{align}
where the states are \emph{cyclic}, and the action of the {closed} chain Hamiltonian is given by
\beq 
H_{\text{closed}} |\overset{k}{\bullet} \overset{J-k}{\bullet}\rangle = H_{12}  |\overset{k}{\bullet} \overset{J-k}{\bullet}\rangle+ H_{12}  |\overset{J-k}{\bullet} \overset{k}{\bullet}\rangle \,.
\eeq
As the states are short, for general $\bar z < \infty$ the $sl(2)$ magnons effectively hop both left and right, and the stampede is given by
\beq
\label{20prime_stampede}
\mathcal{G}(z,\bar z) \simeq \sum_{J=0}^{\infty} \bar z^J  \langle \overset{J}{\bullet} \circ | z^{\lambda H_{\text{closed}}}|\circ \overset{J}{\bullet} \rangle =  \sum_{n=0}^\infty \frac{(\lambda \log(z))^n}{n!}\,\overbrace{\sum_{J=0}^\infty \bar z^J\,{ \langle \overset{J}{\bullet} \circ |(H_\texttt{closed})^n |\circ \overset{J}{\bullet} \rangle}}^{\displaystyle 4^n f_n(\bar z)} \,.
\eeq
The data for the first few loops read
\begin{align}
\begin{aligned}
\label{20prime_data}
&f_0(\bar z) =1 +\frac{1}{1-\bar z} \,,\\&
f_1(\bar z) =\frac{\bar z }{2(\bar z-1)} \textrm{H}_{1}(\bar z) \,,\\
&f_2(\bar z)=\frac{1}{2(1- \bar z)}\left(\bar z\, \textrm{H}_{2}(\bar z)+\textrm{H}_{1,1}(\bar z)\right) \,,&\\
&f_3 (\bar z)=  \frac{2 \bar z\, \textrm{H}_{3}(\bar z)-(\bar z-4)\, \textrm{H}_{1,2}(\bar z)+2\, \textrm{H}_{2,1}(\bar z)+3\,
   \textrm{H}_{1,1,1}(\bar z)}{4( \bar z-1)}\,,\\
&f_4 (\bar z)
   =   -\frac{\bar z\, \textrm{H}_{4}(\bar z)-(\bar z-3) \,\left( \textrm{H}_{1,3}(\bar z)+ \textrm{H}_{2,2}(\bar z)\right)+\textrm{H}_{3,1}(\bar z)}{2 (\bar z-1)}+ \\ &\;\;\;\;\qquad+\frac{3\,\textrm{H}_{1,1,2}(\bar z)+2\,
   \textrm{H}_{1,2,1}(\bar z)+ 2\, \textrm{H}_{2,1,1}(\bar z)+3 \textrm{H}_{1,1,1,1}(\bar z)}{2 (\bar z-1)}\,.
\end{aligned}
\end{align}
The harmonic polylogarithms appearing in \eqref{20prime_data} are either standard polylogs $\Li_k(\bar z)$ or combinations of products thereof, while starting at five loops more general HPLs appear. 
Plugging \eqref{20prime_data} into \eqref{20prime_stampede} one can easily check the match with the OPE \eqref{LC_OPE_light}.
In the double light-cone double-scaling limit defined by
\beq
\lambda \to 0^{-}\,,\,\, z\to 0^{+}\,,\,\,\bar z\to \infty \,,\,\,\,\,\,\, s^2=\lambda \log(z) \log(\bar z)<\infty \,,
\eeq
the correlator of $\mathbf{20'}$ operators is equal to a the square of a null octagon with $l=0$ bridges. Indeed, making use of the asymptotic of $\textrm{H}$s \eqref{HPLasympt} we check that
 \beq
\mathcal{G}(0, \infty) \simeq \mathbb{O}_0(2s) \mathbb{O}_0(2s)  = e^{-{2 s^2}}\,.
\eeq
As a last comment, we point out that a different choice of polarization vectors $y_i$ in \eqref{20p_4pt} would require a different hamiltonian. For example, 
\beq
\langle\Tr\left( X ^2\right)(x_1)\Tr\left(\bar X Z \right)(x_2)\Tr\left( Z \bar X \right)(x_3)\Tr\left(\bar Z^2 \right)(x_4) \rangle\,,
\eeq
is decomposed into hopping states
\beq 
\Tr(X e^{\mathbb K_-} Z)(\infty,\infty)\,,\,\,\,\,\, \text{and}\,\,\,\,\, \Tr(\bar X e^{\bar z \mathbb P_-}\bar Z)(0,0)\,.
\eeq
The presence of fields with different $SU(4)$ flavour $X,Z$ requires to compute the stampede with the $psu(1,1|2)$-chain hamiltonian acting on two sites, as explained in appendix \ref{app:general_octagon}. In fact, the same situation would appear in \eqref{20p_4pt} taking the light-cone limit $x_{13}^2,x_{24}^2\to 0$. 

Why couldn't we just do a similar check for larger operators? A key simplification of the $20'$ correlator is the fact that there is a single leading twist ($=2$) operator for each spin. This single operator has a well known anomalous dimension (\ref{gammaS}) so that we can right away dress the tree level structure constant by powers of this anomalous dimension in the OPE expansion, preform the conformal partial wave sums and that is it. For larger external operators, however, there is \textit{degeneracy}: there will be many operators flowing in the OPE for each given spin. To generate logs we thus need to compute the structure constants for \textit{each} of these operators and dress \textit{each} structure constant by it own anomalous dimension. In other words we have to compute sum rules as explained in \cite{VW}. This advantage of that integrability based approach is that once we lift the degeneracy (by solving all one-loop Bethe equations for a fixed spin) we literally just need to exponentiate the corresponding dimensions to get the leading logs \cite{VW}. The drawback is that the degeneracy grows quickly with the spin and solving Bethe equations for larger and larger spin becomes hard. The stampedes bypass this altogether. Up to the number of loops we checked they work much more efficiently. On the other hand, going beyond leading logs seems unfeasible within the stampede framework, while it is straightforward in the integrability-based computation using hexagons \cite{hexagons,morehexagons1,morehexagons2,morehexagons3}.

\section{Toda in a more general light-cone limit.}
\label{app:toda}
The octagon with zero bridges is amazingly simple in the double light-cone limit \cite{BK0}:
\beq
\mathbb{O}_{0}(z,\bar z) \simeq \exp \left(- \log(z/\bar{z})^2 \frac{\log(\cosh(2\pi g))}{8\pi^2} +\log(z \bar{z})^2  \frac{g^2}{4}- \frac{\log(\sinh(4\pi g)/(4\pi g)) }{8} \right) \la{start0}\,,
\eeq
and in this section we find more convenient to use the coupling $g^2=-\lambda$ in order to avoid to use square roots.
This expression holds for $z$ very large, $\bar z$ small and for any order in perturbation theory. 
When $g\to 0$ we can simplify this expression to 
\beq
\mathbb{O}_{0}(z,\bar z) \simeq \exp \left(- \log(z/\bar{z})^2 \Big(\frac{g^2}{4}-\frac{\pi^2 g^4}{6} \Big) +\log(z \bar{z})^2  \frac{}{4}  \right)
\eeq
which simplifies in two different nice ways when we take a further single or double \textit{light-cone double scaling (LCDS)} limit and hold some products of cross-ratios and coupling fixed while taking the coupling to zero:
\beq
\mathbb{O}_{0}(z,\bar z) \simeq 
\left\{
\begin{array}{ll}
\exp(- L \bar L-\frac{\pi^2}{6} L^2)
&\,\,, L \equiv g^2 \log(z) \,\, , \, \, \bar{L} \equiv g^0 \log(1/\bar{z}) \,\,,\,\, \texttt{single LCDS limit}
\\ \\
\exp({- L \bar L} )
&\,\,, L \equiv g \log(z) \,\, , \, \, \bar{L} \equiv g \log(1/\bar{z}) \,\,\,\,\,\, ,\,\,\texttt{double LCDS limit}
\end{array}
\right. \nn
\eeq
The first limit is clearly more general. It reduces to the second one by taking $L$ small and $\bar L$ large. 

The octagon with a single bridge is already much richer. Before taking any scaling limit the analogue of (\ref{start0}) already reads 
\beqa
\label{tau_2D}
\frac{\mathbb{O}_1 }{ \mathbb{O}_0} &\simeq& {\color{blue}I_0} \label{I0Plus}\\ &+&
\frac{\pi^2}{3} g^2 \left( {\color{blue} (I_2-2 s I_1) \cos (\varphi )}+I_0-2 s I_1\right) \nn \\&+&
\frac{\pi^4}{9} g^4
   \left({\color{blue} \frac{32 s I_3+ (20 s^2+7 ) I_4}{20}   \cos (2 \varphi )} +    ( (4 s^2+9 ) I_2+4 s I_3 ) \cos (\varphi)\right. +\\& &\qquad\qquad\qquad\qquad\qquad\qquad\qquad\qquad\quad+ \left. \frac{ (60 s^2 -43 ) I_0+48 s I_1}{20}\right)\nn + \\
   &+&g^6 \pi^6 \left({\color{blue} -\frac{ \left(\left(280 s^3+2378
   s\right) I_5+\left(364 s^2-93\right) I_6\right) \cos (3 \varphi )}{22680}}+\dots 
   \right)  +\dots \nn
\eeqa
where the argument of all modified Bessel functions is $I_n=I_n(2s)$ and where
\beq
s^2 \equiv g^2 \log(z) \log(1/\bar{z}) \qquad , \qquad e^{i \varphi} \equiv  \frac{\log(z)}{ \log(1/\bar{z})}\,.
\eeq
In the \textit{single} LCDS limit we hold  
\beq
L \equiv g^2 \log(z) \,\, , \qquad \bar{L} \equiv g^0 \log(1/\bar{z})
\eeq
fixed. Then each cosine in (\ref{I0Plus}) blows up and we keep the first term in each line marked in blue. We thus get 
\beq
 \frac{\mathbb{O}_1 }{ \mathbb{O}_0} \simeq I_0  + \left(\frac{L}{\bar L}\right) \frac{\pi ^2}{6 } \left(I_2 -2 s \,I_1 \right)+\left(\frac{L}{\bar L}\right)^2 \frac{\pi ^4}{360} \left(\left(20 s^2+7\right) I_4+32 s I_3\right)+\dots \equiv \tau(L,\bar L) \la{tauEq}
 \eeq
where as above $I_n=I_n(2s)=I_n(2 \sqrt{L \bar L})$. The function $\tau(L,\bar L)$ defined in this way is quite an important function. In the \textit{single} LCDS limit all octagons can be nicely expressed in terms of this same function as 
\beq
 \frac{\mathbb{O}_l }{ \mathbb{O}_0} \simeq  \det_{i,j\le l} \left[
 \Big(\frac{\partial}{\partial L}\Big)^{i-1}  \Big(\frac{\partial}{\partial \bar L}\Big)^{j-1} \tau(L,\bar L)  \label{det}
 \right] \qquad \texttt{\underline{single} LCDS limit} \,.
 \eeq
In other words, the functions $\tau_l\equiv  \frac{\mathbb{O}_l }{ \mathbb{O}_0}$ obey a two-dimensional Toda equation
\beq
\frac{\tau_{l-1} \tau_{l+1} }{\tau_l^2} - \partial_L \partial_{\bar L} \log \tau_l  =0 \,.
\eeq
This had already been observed in the \textit{double} LCDS limit by Belitsky and Korchemsky in~\cite{BK}. In this simpler case all cosines in (\ref{I0Plus}) are finite so only the first line survives and thus  $\tau = \frac{\mathbb{O}_1 }{ \mathbb{O}_0} \simeq I_0$. In this case the determinant in (\ref{det}) can be neatly simplified into a Bessel Henkel determinant
\beq
 \frac{\mathbb{O}_l }{ \mathbb{O}_0} \simeq  \det_{i,j\le l} I_{|i-j|}\qquad \texttt{\underline{double} LCDS limit} \,.
\eeq
Would be interesting to generalize the derivation in \cite{BK} to the \textit{single} LCDS limit to understand why a Toda equation still governs this limit. 

It would also be very nice to find a simple closed expression for $\tau(L,\bar L)$ by resumming~(\ref{tauEq}). Knowing $\tau$ and therefore \eqref{det} one can fix the constants $c_{i} ^{(\ell)}$ defining the octagon \eqref{AsympDeterminant} until order nine loops. This can be done by plugging the DLC expression for the ladders
\begin{equation}
\medmuskip=0.5mu
\thinmuskip=1mu
\thickmuskip=1mu
f_p(z,\bar{z})\,\, \simeq \underset{\substack{m,n=0, \\ m+n\, \text{mod}\, 2 =0 }}{\sum^p} -\frac{(p-1)! p! \left(2-2^{m+n-2 p+2}\right) (-m-n+2 p)! \zeta (-m-n+2 p)}{m! n! (p-m)! (p-n)!} \log(z)^m \log(\bar z)^n \,, 
\end{equation}
inside equation \eqref{AsympDeterminant}, and equating it with the expression $\mathbb{O}_l/\mathbb{O}_0$ in the single LCDS, where on both sides of the equation the expansion is of the type
\beq
1+ s^2 \textit{t}_{1,0} + s^4 ( x^2 \textit{t}_{2,2} + \textit{t}_{2,0}) + s^6 ( x^2 \textit{t}_{3,2} + \textit{t}_{3,0})\dots +s^{2m}  \underset{\substack{k=0, \\ k\,\text{mod}\, 2 =0 }}{\sum^m}  x^k t_{m,k}+\dots\,,
\eeq
for $s^2= g^2 \log(z)\log(1/\bar z)$ and $x = \log(\bar z)$.
At higher orders in the coupling, in the single LCDS limit there are not sufficient equations to disentangle all the constants $c_{i} ^{(\ell)}$. Indeed, while at order $s^{2m}$ there are $m/2$ (or $(m-1)/2$) equations for even (odd) $m$, the number of variables $c_{i} ^{(\ell)}$ is related to the number of partitions entering \eqref{AsympDeterminant}, its growth being faster than linear.  


\section{Octagon Bootstrap}
\label{app:general_octagon}
The method of stampedes can be applied to a wider range of computations of light-cone correlators than the one explained in section \ref{sect:SL2}, which is particularly accessible and instructive as - since all fields involves are of the same type $Z$ - the stampede is governed by the simple $sl(2)$ Hamiltonian. In fact, the general octagon function is a mere generalization of \eqref{AsympDeterminant} by a function $\chi_n(z, \bar z,\alpha, \bar \alpha)$ of conformal and $SU(4)$ cross ratios \cite{Determinant}
 \beq
 \label{gen_chi}
\mathbb{O}_{\chi,l} (z,\bar{z},\alpha,\bar{\alpha}) = 1+ \sum_{n=1}^{\infty} \sum_{m=n(n+l)}^{\infty} \lambda^m \sum_{\underline{k} \in S^{(m,n)}}  \chi_n(z,\bar{z},\alpha,\bar{\alpha})\, c^{(l)}_{k_1,\dots,k_n}{\mathcal{S}}_{k_1,\dots,k_n}(z,\bar{z})\,,
\eeq
and the $SU(2|2)$ character is
\beq
 \chi_n(z,\bar{z},\alpha,\bar{\alpha}) = \frac{\left(-\frac{(z-\alpha)(\bar{z}-\bar{\alpha})}{\alpha}\right)^n+\left(-\frac{(z-\bar{\alpha})(\bar{z}-{\alpha})}{\bar{\alpha}}\right)^n}{2}\,,
 \eeq
 therefore the knowledge of $c_i^{(l)}$ in the \emph{asymptotic} case $\chi_n \to (z+\bar z - z \bar z)^n$ worked out in section \ref{sect:SL2} fixes the most general case.
 \begin{figure}[t]
 \begin{center}
{\includegraphics[scale=0.62]{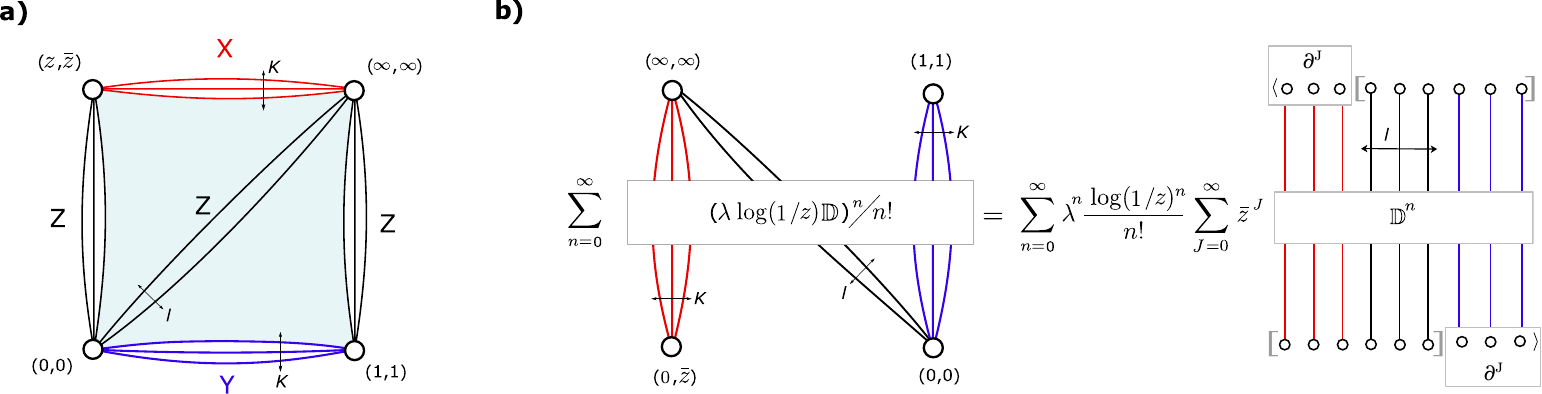}}
 \end{center}
 \caption{\textit{a)} The asymptotic correlator of figure \ref{SL2_asymp_fig} after a change of conformal frame $(z,\bar z) \leftrightarrow (1,1)$. We highlight the interior of the frame, corresponding to an octagon with $l$ bridges. \textit{b)} The $psu(1,1|2)$ stampede computation of the octagon described in formula \eqref{simp_LC_stampede}.}
 \label{gen_oct}
 \end{figure}
It is nevertheless interesting to check that the same set of arguments which lead from a light-cone correlator to the stampede can be applied directly in presence of fields with different flavours (e.g. $X,Y,Z$) in the top/bottom effective states. This case includes for instance the light-cone behaviour of the same correlator of section \ref{sect:SL2} but in the $t$-channel $x_{13}^2,x_{24}^2\to 0$, depicted in figure \ref{gen_oct} or - which in the light cone-limit is the same - of the simplest octagon corresponding to \eqref{gen_chi} with
\beq
\label{chi_symp}
 \chi_n(z,\bar{z},\alpha,\bar{\alpha})\to  \left({-(1-\bar z)(1-z)}\right)^n\,,
\eeq
In this situation the fields $Z$ connecting the edges $x_{13}, x_{24}$ are only spectators, while they enter the computation as bridges. 
For thick edges $K\gg1$ we assume the factorization into octagons, and after a rotation of the frame by $-\pi/2$, the LC expansion of the null edges read
\begin{align}
&O^\text{eff}_\text{bottom} \simeq \![Y^{K}Z^{l}] e^{\bar z \mathbb{P}} X^K (0) =  
\sum_{J=0}^\infty\frac{ \bar z^J }{J!}  |[Y^{K}Z^{l}] {\partial^J} X^K  \rangle\,,\\&O^\text{eff}_\text{top} \simeq \! [\bar X^{K}\bar {Z}^l] e^{\mathbb{K}} \bar Y^K (\infty)  =  
\sum_{J=0}^\infty \frac{1}{J!} \langle\left( {\partial^J}  \bar Y^K \right) [\bar {Z}^l \bar X^{K}]|\,.
\end{align}
and we used the symbol $[ABC]=ABC+\text{permutations}$, as the effective state shall remember that the original operators are symmetric traceless products of fields. The light-cone simplest octagon is capture by a stampede computation
\begin{align}
\begin{aligned}
\label{simp_LC_stampede}
\widetilde{\mathbb{O}}_{l}(z,\bar z) =  \sum_{n=l+1}^{\infty} \frac{(\lambda \log(1/z)) ^n}{n!} \sum_{J=0}^{\infty} (\texttt{open psu(1,1|2) stampede})_{J,l,n} \,\bar z ^J\,,
\end{aligned}
\end{align} 
where the evolution of a ket is governed by the famous one-loop dilatation $\mathbb{D}$ operator of $\mathcal{N}=4$ SYM  \cite{Beisert} 
\beq
 (\texttt{open psu(1,1|2) stampede})_{J,l,n} = {}_{\text{open}} 
\langle \texttt{out} |\mathbb{D}^n | \texttt{in} \rangle_{\text{open}}\,. 
\eeq
We checked with Mathematica that the result coincides with the $sl(2)$ case after a rescaling of coefficients $c^{(l)}_{k_1,\dots,k_n}$ by a simple rational function $(\bar z-1)^n/\bar z^n$. This fact matches the expectation from \eqref{gen_chi}, which connects the asymptotic to the simplest octagon by the replacement of the asymptotic character with the simplest one \eqref{chi_symp}, namely
\beq
(z+\bar z - z \bar z)^n c^{(l)}_{k_1,\dots,k_n} \to (-(1-z)(1-\bar z))^n c^{(l)}_{k_1,\dots,k_n}\,,
\eeq
that on the light cone $z \to 0$ reduces to
\beq
\bar z^n c^{(l)}_{k_1,\dots,k_n} \to (\bar z-1)^n c^{(l)}_{k_1,\dots,k_n}\,.
\eeq
 As it stands, it seems quite non-trivial that a totally different Hamiltonian (the psu(1,1|2) one) reproduces the exact same stampedes as the sl(2) one once we dress these constants by these very simple rational factors. There must be a simple group theory explanation. Would be nice to work it out and make as much symmetry manifest as possible.


\begin{thebibliography}{0}

\bibitem{Frank}
F.~Coronado,
``Perturbative four-point functions in planar $ \mathcal{N}=4 $ SYM from hexagonalization,''
JHEP \textbf{01}, 056 (2019)
[arXiv:1811.00467].

\bibitem{FrankBootstrap}
F.~Coronado,
``Bootstrapping the Simplest Correlator in Planar $\mathcal N = 4$ Supersymmetric Yang-Mills Theory to All Loops,''
Phys. Rev. Lett. \textbf{124} (2020) no.17, 171601
[arXiv:1811.03282].

\bibitem{Frobenius}
G.~Frobenius,
"Uber die charaktere der symmetrischer gruppe,"
Preuss. and ad. Wk. sitz. (1900), 516a534.
%
\bibitem{Hook}
Frame, Robinson, and Thrall,
``The Hook Graphs of the Symmetric Group.'',
Canadian Journal of Mathematics 6 (1954): 316a24
\bibitem{Gessel}
I.M.~Gessel, and G.~ Viennot,
Advances in Mathematics, (1985), 58, 300-321.

\bibitem{BassoDixon}
B.~Basso and L.~J.~Dixon,
``Gluing Ladder Feynman Diagrams into Fishnets,''
Phys. Rev. Lett. \textbf{119}, no.7, 071601 (2017)
[arXiv:1705.03545].

\bibitem{BassoDixonAndFriends}
B.~Basso, L.~J.~Dixon, D.~A.~Kosower, A.~Krajenbrink and D.~l.~Zhong,
``Fishnet four-point integrals: integrable representations and thermodynamic limits,''
JHEP \textbf{07}, 168 (2021)
[arXiv:2105.10514].

\bibitem{FishnetJoao}
J.~Caetano, \"O.~G\"urdo\u{g}an and V.~Kazakov,
``Chiral limit of $ \mathcal{N} $ = 4 SYM and ABJM and integrable Feynman graphs,''
JHEP \textbf{03}, 077 (2018)
[arXiv:1612.05895].

\bibitem{Fishnet}
\"O.~G\"urdo\u{g}an and V.~Kazakov,
``New Integrable 4D Quantum Field Theories from Strongly Deformed Planar $\mathcal N = $ 4 Supersymmetric Yang-Mills Theory,''
Phys. Rev. Lett. \textbf{117}, no.20, 201602 (2016)
[arXiv:1512.06704].

\bibitem{Aitken}
A. C~ Aitken, 
"The monomial expansion of determinantal symmetric functions,"
Proc. Royal Soc. Edinburgh (A) 61 (1943), 300-310.

\bibitem{gen_fish}
D.~Chicherin, V.~Kazakov, F.~Loebbert, D.~M\"uller and D.~l.~Zhong,
Phys. Rev. D \textbf{96}, no.12, 121901 (2017)
doi:10.1103/PhysRevD.96.121901

\bibitem{AldayBissi}
L.~F.~Alday and A.~Bissi,
``Higher-spin correlators,''
JHEP \textbf{10}, 202 (2013)
[arXiv:1305.4604].

\bibitem{Beisert}
N.~Beisert,
``The complete one-loop dilatation operator of N=4 superYang-Mills theory,''
Nucl. Phys. B \textbf{676}, 3-42 (2004)
[arXiv:hep-th/0307015].

\bibitem{Asymp_4pt}
B.~Basso, F.~Coronado, S.~Komatsu, H.~T.~Lam, P.~Vieira and D.~l.~Zhong,
``Asymptotic Four Point Functions,''
JHEP \textbf{07}, 082 (2019)
[arXiv:1701.04462].

\bibitem{VW}
P.~Vieira and T.~Wang,
JHEP \textbf{10}, 035 (2014)
[arXiv:1311.6404].
\bibitem{SVW}
A.~Sever, P.~Vieira and T.~Wang,
JHEP \textbf{12}, 065 (2012)
[arXiv:1208.0841].

\bibitem{StrongOctagon}
T.~Bargheer, F.~Coronado and P.~Vieira,
``Octagons II: Strong Coupling,''
[arXiv:1909.04077].

\bibitem{Alday:2010zy}
L.~F.~Alday, B.~Eden, G.~P.~Korchemsky, J.~Maldacena and E.~Sokatchev,
``From correlation functions to Wilson loops,''
JHEP \textbf{09} (2011), 123
[arXiv:1007.3243].

\bibitem{BK0}
A.~V.~Belitsky and G.~P.~Korchemsky,
``Exact null octagon,''
JHEP \textbf{05}, 070 (2020)
doi:10.1007/JHEP05(2020)070
[arXiv:1907.13131].

\bibitem{BGV}
P.~Vieira, V.~Goncalves and C.~Bercini,
``Multipoint Bootstrap I: Light-Cone Snowflake OPE and the WL Origin,''
[arXiv:2008.10407].

\bibitem{BGHV}
C.~Bercini, V.~Gon\c{c}alves, A.~Homrich and P.~Vieira,
``The Wilson Loop - Large Spin OPE Dictionary,''
[arXiv:2110.04364]

\bibitem{BK}
A.~V.~Belitsky and G.~P.~Korchemsky,
``Crossing bridges with strong Szego limit theorem,''
[arXiv:2006.01831].

\bibitem{Determinant}
I.~Kostov, V.~B.~Petkova and D.~Serban,
``Determinant Formula for the Octagon Form Factor in $N$=4 Supersymmetric Yang-Mills Theory,''
Phys. Rev. Lett. \textbf{122}, no.23, 231601 (2019)
[arXiv:1903.05038].

\bibitem{Derkachov:2019tzo}
S.~Derkachov and E.~Olivucci,
``Exactly solvable magnet of conformal spins in four dimensions,''
Phys. Rev. Lett. \textbf{125}, no.3, 031603 (2020)
[arXiv:1912.07588].

\bibitem{Derkachov:2020zvv}
S.~Derkachov and E.~Olivucci,
``Exactly solvable single-trace four point correlators in $\chi$CFT$_4$,''
JHEP \textbf{02}, 146 (2021)


\bibitem{Derkachov:2021rrf}
S.~Derkachov and E.~Olivucci,
``Conformal quantum mechanics \& the integrable spinning Fishnet,''
[arXiv:2103.01940].

\bibitem{Olivucci:2021cfy}
E.~Olivucci,
``Hexagonalization of Fishnet integrals I: mirror excitations,''
[arXiv:2107.13035].

\bibitem{Grabiner}
 D. J.~Grabiner and P. M.~Magyar,
 "Random Walks in Weyl Chambers and the Decomposition of Tensor Powers"
Journal of Algebraic Combinatorics, \textbf{239-260}, 2, (1993).

\bibitem{GZ}
Gessel, I. M.,  Zeilberger, D. (1992). "Random walk in a Weyl chamber." Proceedings of the American Mathematical Society, 115(1), 27-31.

\bibitem{OkounkovReshetikhin}
A.~Okounkov, Andrei, and N.~Reshetikhin,
"Correlation function of Schur process with application to local geometry of a random 3-dimensional Young diagram.",
Journal of the American Mathematical Society 16.3 (2003): 581-603.

\bibitem{LeDouss}
B.~Lacroix-A-Chez-Toine, P.~Le Doussal, S.N.~Majumdar, G.~Schehr,
 "Non-interacting fermions in hard-edge potentials." ,
 Journal of Statistical Mechanics: Theory and Experiment 2018.12 (2018): 123103.

\bibitem{Johnasson}
K.~Johansson,
 "Non-intersecting paths, random tilings and random matrices." 
Probability theory and related fields 123.2 (2002): 225-280.

\bibitem{Baik}
J.~Baik, 
"Random vicious walks and random matrices.",
 Communications on Pure and Applied Mathematics: A Journal Issued by the Courant Institute of Mathematical Sciences 53.11 (2000): 1385-1410.

\bibitem{MNS}
K.~Johansson,
"From Gumbel to Tracy-Widom.",
Probability theory and related fields 138.1 (2007): 75-112.

\bibitem{FrankSimon}
S.~Caron-Huot and F.~Coronado,
``Ten dimensional symmetry of $N$ = 4 SYM correlators,''
[arXiv:2106.03892].

\bibitem{MiguelEtAl}
A.~Antunes, M.~S.~Costa, V.~Goncalves and J.~V.~Boas,
``Lightcone Bootstrap at higher points,''
[arXiv:2111.05453].

\bibitem{decagons}
T.~Fleury and V.~Goncalves,
``Decagon at Two Loops,''
JHEP \textbf{07}, 030 (2020)
[arXiv:2004.10867].

\bibitem{hexagons}
B.~Basso, S.~Komatsu and P.~Vieira,
``Structure Constants and Integrable Bootstrap in Planar N=4 SYM Theory,''
[arXiv:1505.06745].

\bibitem{Hexagonalization}
T.~Fleury and S.~Komatsu,
``Hexagonalization of Correlation Functions,''
JHEP \textbf{01}, 130 (2017)
[arXiv:1611.05577].

\bibitem{HPL}
E.~Remiddi and J.~A.~M.~Vermaseren,
``Harmonic polylogarithms,''
Int. J. Mod. Phys. A \textbf{15}, 725-754 (2000)
[arXiv:hep-ph/9905237 [hep-ph]].

\bibitem{DolanOsb}
F.~A.~Dolan and H.~Osborn,
``Conformal partial wave expansions for N=4 chiral four point functions,''
Annals Phys. \textbf{321}, 581-626 (2006)
[arXiv:hep-th/0412335].

\bibitem{morehexagons1}
B.~Basso, F.~Coronado, S.~Komatsu, H.~T.~Lam, P.~Vieira and D.~l.~Zhong,
``Asymptotic Four Point Functions,''
JHEP \textbf{07} (2019), 082
[arXiv:1701.04462].

\bibitem{morehexagons2}
B.~Eden and A.~Sfondrini,
``Three-point functions in ${\cal N}=4$ SYM: the hexagon proposal at three loops,''
JHEP \textbf{02} (2016), 165
[arXiv:1510.01242]. $\bullet$
B.~Basso, V.~Goncalves, S.~Komatsu and P.~Vieira,
``Gluing Hexagons at Three Loops,''
Nucl. Phys. B \textbf{907} (2016), 695-716
[arXiv:1510.01683]. 

\bibitem{morehexagons3}
B.~Basso, V.~Goncalves and S.~Komatsu,
``Structure constants at wrapping order,''
JHEP \textbf{05} (2017), 124
[arXiv:1702.02154].


\end{thebibliography}
\end{document}